\newcommand\apjcls{1}
\newcommand\aastexcls{2}
\newcommand\othercls{3}

\newcommand\papercls{\aastexcls}
\documentclass[tighten, times, twocolumn]{aastex62}  

\if\papercls \apjcls
\usepackage{apjfonts}
\else\if\papercls \othercls
\usepackage{epsfig}
\usepackage{margin}
\usepackage{times}
\fi\fi
\usepackage{ifthen}
\usepackage{natbib}
\usepackage{amssymb, amsmath}

\if\papercls \apjcls
\newcommand\aas{\ref@jnl{AAS Meeting Abstracts}}
\newcommand\dps{\ref@jnl{AAS/DPS Meeting Abstracts}}
\newcommand\maps{\ref@jnl{MAPS}}
\else\if\papercls \othercls
\usepackage{astjnlabbrev-jh}
\fi\fi

\if\papercls \aastexcls
\hypersetup{citecolor=blue, 
            linkcolor=blue, 
            menucolor=blue, 
            urlcolor=blue}  
\else
\usepackage[
bookmarks=true,           
colorlinks=true,          
citecolor=blue,           
linkcolor=blue,           
menucolor=blue,           
urlcolor=blue,            
linkbordercolor={0 0 1},  
pdfborder={0 0 1},
frenchlinks=true]{hyperref}
\fi

\makeatletter
\DeclareRobustCommand{\lowcase}[1]{\@lowcase#1\@nil}
\def\@lowcase#1\@nil{\if\relax#1\relax\else\MakeLowercase{#1}\fi}
\makeatother

\DeclareSymbolFont{UPM}{U}{eur}{m}{n}
\DeclareMathSymbol{\umu}{0}{UPM}{"16}
\let\oldumu=\umu
\renewcommand\umu{\ifmmode\oldumu\else\math{\oldumu}\fi}

\if\papercls \othercls

\else

\fi

\let\oldsim=\sim
\renewcommand\sim{\ifmmode\oldsim\else\math{\oldsim}\fi}
\let\oldpm=\pm
\renewcommand\pm{\ifmmode\oldpm\else\math{\oldpm}\fi}
\newcommand\by{\ifmmode\times\else\math{\times}\fi}

\newbox{\wdbox}
\renewcommand\c{\setbox\wdbox=\hbox{,}\hspace{\wd\wdbox}}
\renewcommand\i{\setbox\wdbox=\hbox{i}\hspace{\wd\wdbox}}

\newcount\timect
\newcount\hourct
\newcount\minct
\newcommand\now{\timect=\time \divide\timect by 60
         \hourct=\timect \multiply\hourct by 60
         \minct=\time \advance\minct by -\hourct
         \number\timect:\ifnum \minct < 10 0\fi\number\minct}

\catcode`@=11

\newcommand\comment[1]{}

\newcommand\commenton{\catcode`\%=14}

\renewcommand\math[1]{$#1$}
\newcommand\mathshifton{\catcode`\$=3}

\let\atab=&
\newcommand\atabon{\catcode`\&=4}

\let\oldmsp=\sp
\let\oldmsb=\sb
\def\sp#1{\ifmmode
           \oldmsp{#1}%
         \else\strut\raise.85ex\hbox{\scriptsize #1}\fi}
\def\sb#1{\ifmmode
           \oldmsb{#1}%
         \else\strut\raise-.54ex\hbox{\scriptsize #1}\fi}
\newbox\@sp
\newbox\@sb
\def\sbp#1#2{\ifmmode%
           \oldmsb{#1}\oldmsp{#2}%
         \else
           \setbox\@sb=\hbox{\sb{#1}}%
           \setbox\@sp=\hbox{\sp{#2}}%
           \rlap{\copy\@sb}\copy\@sp
           \ifdim \wd\@sb >\wd\@sp
             \hskip -\wd\@sp \hskip \wd\@sb
           \fi
        \fi}
\def\msp#1{\ifmmode
           \oldmsp{#1}
         \else \math{\oldmsp{#1}}\fi}
\def\msb#1{\ifmmode
           \oldmsb{#1}
         \else \math{\oldmsb{#1}}\fi}

\def\supon{\catcode`\^=7}

\def\subon{\catcode`\_=8}

\def\supsubon{\supon \subon}

\newcommand\actcharon{\catcode`\~=13}

\newcommand\paramon{\catcode`\#=6}

\comment{And now to turn us totally on and off...}

\newcommand\reservedcharson{ \commenton  \mathshifton  \atabon  \supsubon 
                             \actcharon  \paramon}

\catcode`@=12
\reservedcharson

\if\papercls \apjcls

\else

\fi

\newcommand\chisq{\ifmmode{\chi\sp{2}}\else\math{\chi\sp{2}}\fi}
\newcommand\redchisq{\ifmmode{ \chi\sp{2}\sb{\rm red}}
                    \else\math{\chi\sp{2}\sb{\rm red}}\fi}
\newcommand\Teq{\ifmmode{T\sb{\rm eq}}\else$T$\sb{eq}\fi}
\newcommand\mjup{\ifmmode{M\sb{\rm Jup}}\else$M$\sb{Jup}\fi}
\newcommand\rjup{\ifmmode{R\sb{\rm Jup}}\else$R$\sb{Jup}\fi}
\newcommand\msun{\ifmmode{M\sb{\odot}}\else$M\sb{\odot}$\fi}
\newcommand\rsun{\ifmmode{R\sb{\odot}}\else$R\sb{\odot}$\fi}
\newcommand\mearth{\ifmmode{M\sb{\oplus}}\else$M\sb{\oplus}$\fi}
\newcommand\rearth{\ifmmode{R\sb{\oplus}}\else$R\sb{\oplus}$\fi}

\shorttitle{The Impact of Enhanced Halo Resolution on the Simulated CGM}
\shortauthors{Hummels {\em et al.}}

\begin{document}

\title{The Impact of Enhanced Halo Resolution on the\\Simulated Circumgalactic Medium}


\author[0000-0002-3817-8133]{Cameron~B.~Hummels}
\affiliation{TAPIR, California Institute of Technology, Pasadena, CA 91125, USA}
\affiliation{NSF Astronomy and Astrophysics Postdoctoral Fellow}
\author[0000-0002-6804-630X]{Britton~D.~Smith}
\affiliation{San Diego Supercomputer Center, University of California, San Diego, CA 92121, USA}
\author[0000-0003-3729-1684]{Philip~F.~Hopkins}
\affiliation{TAPIR, California Institute of Technology, Pasadena, CA 91125, USA}
\author[0000-0002-2786-0348]{Brian~W.~O'Shea}
\affiliation{Department of Computational Mathematics, Science and Engineering, Michigan State University, East Lansing, MI 48824, USA}
\affiliation{Department of Physics and Astronomy, Michigan State University, East Lansing, MI 48824, USA}
\author[0000-0002-4109-9313]{Devin~W.~Silvia}
\affiliation{Department of Computational Mathematics, Science and Engineering, Michigan State University, East Lansing, MI 48824, USA}
\author[0000-0002-0355-0134]{Jessica~K.~Werk}
\affil{Department of Astronomy, University of Washington, Seattle, WA 98195, USA}
\author[0000-0001-9158-0829]{Nicolas~Lehner}
\affil{Department of Physics, University of Notre Dame, Notre Dame, IN 46556, USA}
\author[0000-0003-1173-8847]{John~H.~Wise}
\affil{Center for Relativistic Astrophysics, Georgia Institute of Technology, Atlanta, GA 30332, USA}
\author[0000-0001-6661-2243]{David~C.~Collins}
\affil{Department of Physics, Florida State University, Tallahassee, FL 32306, USA}
\author[0000-0003-1257-5007]{Iryna~S.~Butsky}
\affil{Department of Astronomy, University of Washington, Seattle, WA 98195, USA}

\email{chummels@gmail.com}


\begin{abstract}
Traditional cosmological hydrodynamics simulations fail to spatially resolve the circumgalatic medium (CGM), the reservoir of tenuous gas surrounding a galaxy and extending to its virial radius.  We introduce the technique of Enhanced Halo Resolution (EHR), enabling more realistic physical modeling of the simulated CGM by consistently forcing gas refinement to smaller scales throughout the virial halo of a simulated galaxy.  We investigate the effects of EHR in the \textsc{tempest} simulations, a suite of \textsc{enzo}-based cosmological zoom simulations following the evolution of an L* galaxy, resolving spatial scales of 500 comoving pc out to 100 comoving kpc in galactocentric radius.  Among its many effects, EHR (1) changes the thermal balance of the CGM, increasing its cool gas content and decreasing its warm/hot gas content; (2) preserves cool gas structures for longer periods; and (3) enables these cool clouds to exist at progressively smaller size scales.  Observationally, this results in a boost in ``low ions'' like \ion{H}{1} and a drop in ``high ions'' like \ion{O}{6} throughout the CGM.  These effects of EHR do not converge in the \textsc{tempest} simulations, but extrapolating these trends suggests that the CGM in reality is a mist consisting of ubiquitous, small, long-lived, cool clouds suspended in a hot medium at the virial temperature of the halo.  Additionally, we explore the physical mechanisms to explain why EHR produces the above effects, proposing that it works both by (1) better sampling the distribution of CGM phases enabling runaway cooling in the denser, cooler tail of the phase distribution; and (2) preventing cool gas clouds from artificially mixing with the ambient hot halo and evaporating.  Evidence is found for both EHR mechanisms occurring in the \textsc{tempest} simulations.
\end{abstract}

\keywords{methods: cosmology: theory}

\section{Introduction}
\label{sec:intro}

\subsection{Observations Indicate Substantial Cool Gas in Halo}

The circumgalactic medium (CGM) is the low-density, multiphase gas surrounding a galaxy and extending to its virial radius and beyond.  The CGM is increasingly recognized for its significant role driving the evolution of galaxies, operating as both the reservoir of gas providing fuel to the galaxy, as well as the sink into which stars and AGN deposit energy, mass, and metals \citep{tumlinson:2017}.

Due to its low-density state, the CGM is most efficiently observed through absorption-line spectroscopy, which has revealed it to be multiphase \citep[e.g.,][]{lanzetta:1995}.  Observations indicate the presence of cool 10$^4 {\rm K}$ CGM gas bearing neutral hydrogen and low ionization-potential ions (``low ions'') like \ion{Mg}{2} and \ion{Si}{2} \citep[e.g.,][]{churchill:1996, chen:1998, steidel:2010, gauthier:2010, matejek:2012, tumlinson:2013, prochaska:2014, werk:2014, johnson:2017}, as well as warm-hot gas (10$^{5.5}$ - 10$^6 {\rm K}$) traced by ``high ions'' in the form of \ion{N}{5}, \ion{O}{6}, and \ion{Ne}{8} \citep[e.g.,][]{stocke:2006, tumlinson:2011, savage:2011, tripp:2011, meiring:2013, pachat:2017, burchett:2018}.  Observational studies consistently demonstrate large quantities and covering fractions for \ion{H}{1} and other low ions in the CGM over many redshifts, environments, and halo masses (e.g., \citealt{rudie:2012}, \citealt{werk:2013}, \citealt{borthakur:2016}).

\subsection{Simulations Underpredict Cool Gas Content in Halo}

Hydrodynamical simulations provide the theoretical groundwork to not only reproduce observational studies of galaxies, but to better understand the processes responsible for galactic evolution.  Numerical galaxy studies have made significant advances in our understanding of: large scale structure formation \citep[e.g.,][]{springel:2005}, the stellar mass function \citep[e.g.,][]{torrey:2014}, the galaxy core-cusp problem \citep[e.g.,][]{pontzen:2012}, the Tully-Fisher Relation \citep[e.g.,][]{schaye:2015}, the Mass-Metallicity Relation \citep[e.g.,][]{dave:2011}, and more (see \citealt{somerville:2015} for a full listing).  However, simulations struggle to reproduce the observational characteristics of the CGM.  Specifically, simulations chronically underproduce the observed column densities of various ions observed to be present in galactic halos.  The low abundances of CGM ions in simulations have been demonstrated for grid-based codes \citep{hummels:2013, liang:2016}, particle-based codes \citep{stinson:2012, shen:2012}, and moving-mesh codes \citep{suresh:2015}; for low redshift \citep{ford:2013}, high redshift \citep{bird:2016}, and idealized simulations \citep{fielding:2017a}; and affecting both low ions \citep{fumagalli:2011, oppenheimer:2018a} and high ions \citep{roca:2018}.  This underproduction of ionic absorbers in simulated galactic halos is a major problem for our understanding of the CGM and increasingly for galaxy evolution as a whole.

\begin{figure}
\centering
\includegraphics[width=1.0\columnwidth, clip]{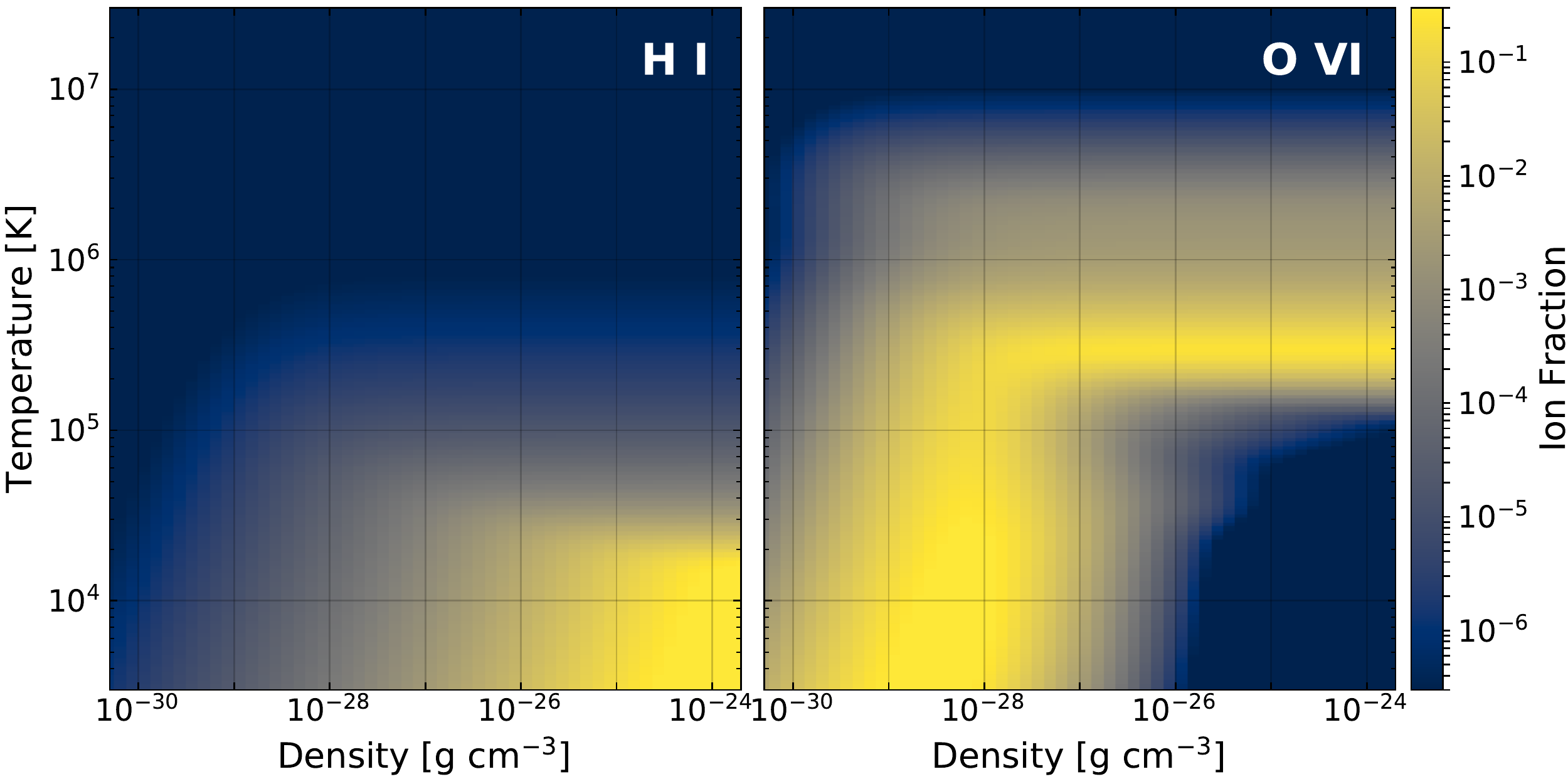}
\caption{Ion fractions of \ion{H}{1} and \ion{O}{6} as a function of density and temperature.  \ion{H}{1} and low ions probe very different gas phases than \ion{O}{6} and other high ions, so recent methods successful at increasing simulated \ion{O}{6} CGM content to observable levels have little impact on addressing the lack of \ion{H}{1} in simulations of the CGM.  These data are taken from data tables included in the \textsc{trident} code \citep{hummels:2017} assuming collisional ionization and photoionization at $z=1$ from a \citet{haardt:2012} UV background.}
\label{fig:ion_fractions}
\end{figure}

Recent work has focused on explaining the deficit of \ion{O}{6}-bearing material in simulated halos compared to the $N_{\rm OVI}$ measurements from the COS-Halos survey (log $N_{\rm OVI}$ $\gtrsim$ 14.0) at $z\sim 0.25$ \citep{tumlinson:2011}, where groups have demonstrated how the inclusion of AGN feedback can increase simulated \ion{O}{6} column densities to observed levels \citep{oppenheimer:2018a, nelson:2018, sanchez:2018}.  However, these successes have had little impact on increasing halo low ion content to reproduce observations.  This is not surprising since the physical properties of the gas hosting low ions is markedly different than the gas hosting high ions.  Figure \ref{fig:ion_fractions} shows the ionization fraction of \ion{H}{1} and \ion{O}{6} across a range of temperatures and densities taken from the data tables in the \textsc{trident} analysis code \citep{hummels:2017}.  \ion{H}{1} tends to reside in cool, dense gas whereas \ion{O}{6} is more abundant in warm, dense gas (collisionally ionized) and cool, rarified gas (photoionized).  Thus, the methods successful at reproducing the high ions do not guarantee success at reproducing the low ions because they probe different phases of gas.

\subsection{Evidence for Cool Gas On Small Scales}

We propose that the primary reason existing simulations have been unable to reproduce the observed column density of low ions is a lack of spatial resolution in the simulated halo. When a simulation resolution element is larger than the natural size scale for a cool gas cloud, the cool gas content artificially mixes with the surrounding hot gas medium, and a spurious warm gas is formed.  Put another way, in the absence of sufficient spatial resolution to resolve the natural gas cloud size, multiphase gas structure is suppressed.  As we will show, the result is too little cool gas, which can lead to the simulations underpredicting the low ions that trace it (e.g., \ion{H}{1}).

There is reason to believe that the size of the low-ion-bearing clouds in the CGM is small ($ \lesssim $ kpc) from observations.  \citet{rauch:1999} found evidence for spatially and kinematically distinct components of \ion{Si}{2} and \ion{C}{2} at $z=3.5$ with separations estimated at $26/h$ pc.  \citet{lehner:2013} show for 28 Lyman Limit Systems (LLSs, $N_{\rm HI} > 10^{17}$ cm$^{-2}$) at $z<1$ that absorber sizes range from a few pc to kpc scales. Lehner et al. (2018, in prep.) find similar results with a sample of 263 absorbers with log $N_{\rm HI}$ = [15.2, 19].  Using careful ionization modeling of the conditions of a $z=2.5$ LLS, \citet{crighton:2015} concluded that the \ion{H}{1}-bearing clouds were $<$ 500 pc in size.  However, \citet{rubin:2018} recently found that \ion{Mg}{2} absorbers have coherence scales at $> 1.9$ kpc. 
\citet{stern:2016} modeled the COS-Halos observations using a universal density profile and showed that low ions have a typical size of 10 - 100 pc, whereas high ions like \ion{O}{6} span tens of kpc.

On the theory side, work has recently been done updating the classical theory of thermal instability \citep{field:1965}.  \citet{mccourt:2018} used extremely high-resolution idealized hydrodynamic simulations to demonstrate how thermally unstable gas ``shatters'' to form cool gas clouds at the sub-parsec scale.  Additionally, ``cloud-crushing simulations'' have been used  to approximate the conditions of the CGM in high resolution (sub-pc) to investigate survival of cool gas clouds in a hot medium.  Many of these simulations show evidence for increased cool gas content under high-resolution conditions \citep{gronke:2018} and by accounting for additional physical effects like magnetic fields \citep{ji:2018a, liang:2018b}, thermal conduction \citep{armillotta:2017}, and hydrodynamic shielding \citep{forbes:2018}. 

\citet{thompson:2016} propose that efficient cooling in resolved, supernovae-driven outflows can produce substantial cool gas content in galactic winds and the CGM, and this is confirmed with extremely high-resolution isolated galaxy simulations at 5 pc spatial scales \citep{schneider:2018b}.  Very recent work further advances the idea that increased spatial resolution results in variable increases in the low-ion content of the CGM \citep{vandevoort:2018, peeples:2018, suresh:2018}.

\subsection{Traditional Simulations Underresolve the CGM}

Traditional numerical studies of the CGM, both cosmological survey and zoom simulations, have resolution elements in the galactic halo that are typically many kpcs, orders of magnitude larger than the proposed cool-gas scales.  At these resolutions they underresolve the CGM, leading to a lack of multiphase structure unless other steps are taken to enhance the resolution in the galactic halo.  Both particle-based and most grid-based simulations under-resolve gas in low-density regions like the CGM, since both techniques were developed to focus their computational power on collapsed structures of high densities like the star-forming galactic disk.  

In a particle-based simulation, including smoothed particle hydrodynamics (SPH) as well as other variants (e.g., \textsc{fire2} simulations -- \citealt{hopkins:2018}), spatial resolution is tied to gas particle density.  Thus, regions of low density, like the galactic halo, will intrinsically lack spatial resolution relative to the galactic disk.  For example, the smoothing length (i.e., spatial resolution) found in the highest resolution \textsc{fire2} simulation \citep{wetzel:2016} achieves an impressive $\sim10$ pc spatial resolution in the disk of the galaxy but only $\sim7$ kpc spatial resolution at the virial radius of a Milky-Way-like galaxy at $z=0$.  Large-volume, non-zoom simulations like \textsc{eagle} \citep{schaye:2015} and \textsc{illustris} \citep{genel:2014} trade off resolution for a larger simulation volume, so it can only be expected that their halo spatial resolutions will be even coarser.

\begin{figure}
\centering
\includegraphics[width=1.0\columnwidth, clip]{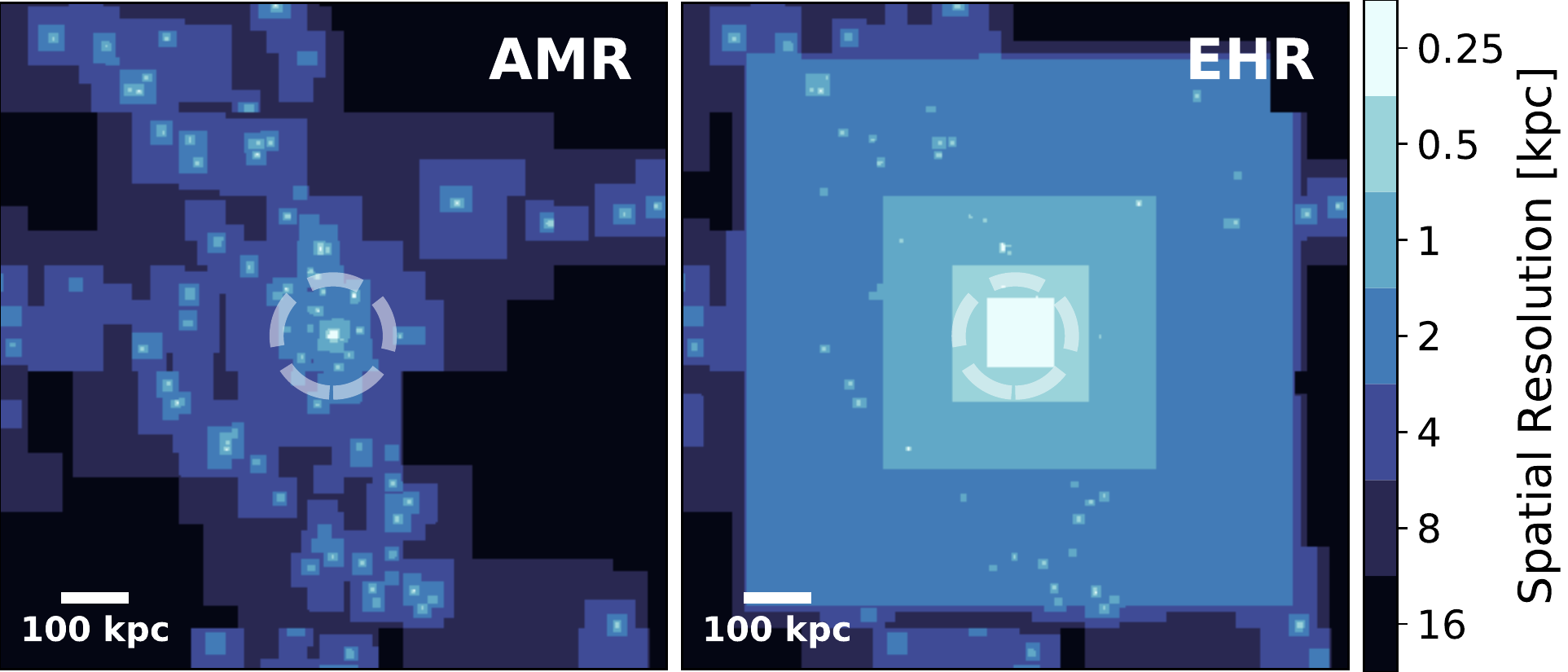}
\caption{Spatial resolution from a traditional AMR simulation (\textbf{left}) versus an Enhanced Halo Resolution (EHR -- 500 co-moving pc) simulation (\textbf{right}).  Plot was generated from a 1-Mpc-sized region at $z=1$ around an L* galaxy from the \textsc{tempest} simulations with the dashed circle representing the virial radius of the galaxy.  Clearly visible are the many galaxies tracing along intergalactic filaments feeding the target galaxy.  EHR works in concert with AMR to assure a base level of spatial resolution of the target galaxy's CGM as well as any gas that will eventually merge with it.}
\label{fig:refinement}
\end{figure}

Grid-based simulations employing adaptive mesh refinement (AMR) trigger increased spatial resolution based on a refinement criterion that is almost universally tied to mass density.  Again, like particle-based simulations, the maximum spatial resolution quoted for such a simulation will be orders of magnitude better than the resolution actually achieved in the low-density galactic halo.  Figure \ref{fig:refinement} demonstrates this, where a traditional cosmological AMR simulation of an L* galaxy achieves 250 pc spatial resolution in the disk of a galaxy but only reaches 4 kpc spatial resolution at the virial radius, a difference of a factor of 16.

\subsection{This Paper: A New Method for Resolving the CGM}

In this work, we introduce the method of Enhanced Halo Resolution, forcing the simulation to achieve a minimum resolution in a fixed region surrounding the target galaxy.  By better resolving a galaxy's halo, it allows the gas to more effectively become multiphase and increases the abundance of cool gas bearing low ions.  The result brings the low-ion content more in line with observations, as well as potentially enables a number of other improvements to the modeling of the CGM.  While the effects of increased resolution do not appear to converge in the simulations in this work, they demonstrate several trends in CGM behavior that can be extrapolated to the predicted convergent parsec scale: A CGM consisting of many pc-scale, long-lived cool gas clouds embedded in a hot halo.  Perhaps more importantly, we describe the physical mechanisms explaining why EHR produces these significant effects in the gaseous halos of galaxies.

Furthermore, we introduce the \textsc{tempest} simulations, a set of cosmological hydrodynamics simulations demonstrating the impact of EHR for an L* galaxy.  The \textsc{tempest} simulations are AMR grid-based simulations run with the \textsc{enzo} code \citep{bryan:2014} for an L* galaxy run from $z=100$ to $z=1$.  Each iteration of the simulation increases the minimum resolution in the halo to address the effects that spatial resolution have on the physical and observed properties of the simulated CGM.  

The layout of our paper is as follows.  In Section \ref{sec:method}, we describe the EHR technique and how it was encoded into the \textsc{tempest} simulations.  Section \ref{sec:discussion} details the various effects that EHR has on the CGM, whereas Section \ref{sec:why} investigates the physical mechanisms in EHR responsible for these effects.  Additional discussion on these results, including predicted convergence of EHR as well as caveats to this study, can be found in Section \ref{sec:caveats}.  Finally, Section \ref{sec:conclusions} lists our conclusions.

\section{Method}
\label{sec:method}

\subsection{Enhanced Halo Resolution}

Traditional simulations achieve spatial resolutions in the low-density galactic halo that are orders of magnitude worse than the scales found in the high-density disk of the galaxy.  This problem cannot be addressed by simply increasing the resolution everywhere in the simulation (e.g., by decreasing mass of gas particles in particle-based codes, or by increasing the base-level resolution in grid-based codes), as it would be prohibitively expensive to do, and would only marginally increase resolution outside of extant high-density structures.  

To address this problem, we introduce the notion of Enhanced Halo Resolution to work in tandem with a traditional density-based adaptive mesh refinement criterion.  Broadly speaking, EHR is any technique to increase resolution in the galactic halo.  In our implementation, we do so by placing a set of nested regions of additional refinement (i.e., ``must-refine regions'') on the center of the galaxy in a simulation to assure that resolution is minimally met out to a fixed galactocentric radius throughout its evolution (See Figure \ref{fig:refinement}).  Each nested region sets a minimum spatial scale that the gas must always achieve.  Each additional region is twice the size of its predecessor and a factor of two coarser in spatial resolution. In our implementation, we chose to use boxes as our nested refined regions, because we found that cubes had the best scaling with our patch-based grid code, \textsc{enzo}.

It is notable that increasing the spatial resolution of the simulation in the outskirts of the galaxy does not introduce any additional physics, subgrid models, or changes in the feedback models that are already successful at reproducing other observational constraints on galactic structure.  Enhancing the halo resolution of the simulation simply results in a more accurate treatment of the physics of the galactic halo on increasingly smaller scales.

The nested nature of the ``must-refine'' regions has two important effects.  One, it assures that a gas cloud traveling toward the galaxy will not cross multiple resolution boundaries in a short period, minimizing any unphysical effects as the cloud passes this artificial barrier between regions of different minimum spatial scale.  Two, the use of nested boxes with powers-of-two differences in \emph{comoving} size/resolution implicitly enables the simulation to achieve a fixed \emph{physical} resolution out to a fixed \emph{physical} galacto-centric radius accounting for the growth of each comoving region due to the expansion of the universe.  

Recall that at redshift $z$, the scale factor of the Universe is smaller by a factor of $a = \frac{1}{1+z}$.  For example, if a simulation uses a single enhanced resolution box covering the galaxy out to a radius of 100 comoving kpc and forcing spatial refinement to 1 comoving kpc spatial resolution, it follows that at redshift $z$, the box only covers the region $\frac{100}{z+1}$ physical kpc at a minimal spatial resolution of $\frac{1}{z+1}$ physical kpc, potentially much smaller than the desired region in physical coordinates.  By nesting several boxes, each meant to be the size of the desired region in physical units at a different redshift, one can assure a base spatial scale in physical units is always met over the cosmic evolution of the galaxy.

The number of nested resolution boxes that one must include in the simulation to consistently meet these physical resolution requirements is dependent on the redshift when EHR is first employed, $z_{\rm start}$.  From that point on, one requires $n_{\rm box}$ nested boxes each a factor of two larger and coarser resolution according to:

\begin{equation}
    n_{\rm box} = \lceil {\rm log}_{\rm 2}{(z_{\rm start}+1)}+1 \rceil
\end{equation}

where $\lceil$ and $\rceil$ denote the ceiling value to assure an integer number of boxes arises.  Thus, to achieve a minimum spatial scale of 1 physical kpc out to 100 physical kpc as far back as redshift 7, when the scale factor of the universe was 0.125, we require 4 nested boxes as shown in Figure \ref{fig:refinement}.

In order to place the refined boxes on top of the galaxy, one must know the center of the galaxy at any given time in the simulation.   This can be done in one of two ways.  The simulation can periodically run an in-line halo finder (e.g., AMIGA, Rockstar, HOP) to determine the centroid of the target galaxy and then pass its centroid to the EHR mechanism as the simulations progresses.  Alternatively, one can first run a traditional AMR version of the simulation to completion, employ a halo finder code to identify the halo centroid over the simulation duration, and then rerun the simulation again with EHR centered on the target galaxy.  In all of our tests, we found the addition of EHR did not change the centroid of our galaxy by more than 1 kpc, making this second method feasible and more easily implemented.

\begin{figure*}
\centering
\includegraphics[width=2.0\columnwidth, clip]{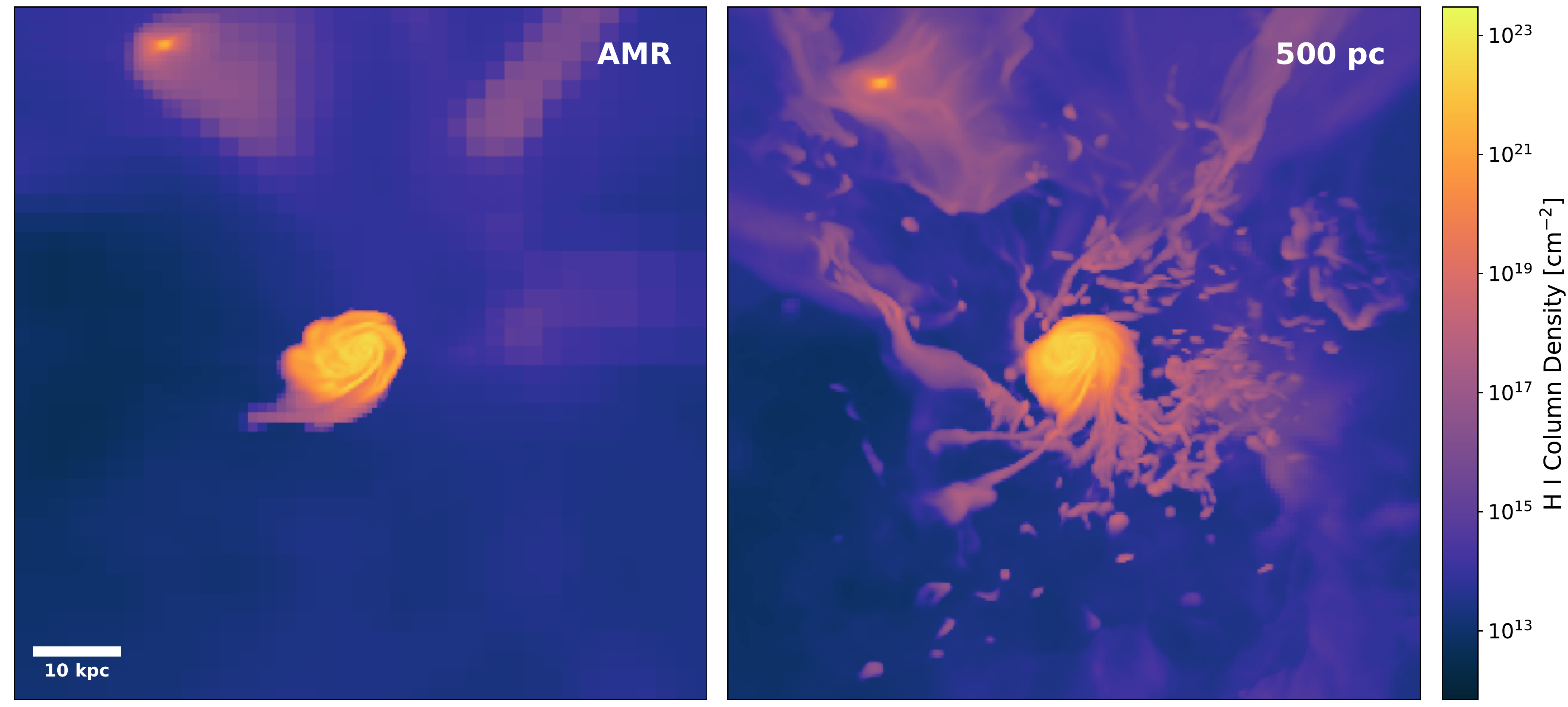}
\caption{\ion{H}{1} column density projections for a simulated galaxy from the \textsc{tempest} simulations at standard AMR resolution versus EHR with 500 co-moving parsecs of CGM resolution.  Images span an 80-kpc-wide region centered on an L* galaxy at $z=1$ showing the inner half of the virial halo.  EHR results in significant changes including an increase in \ion{H}{1} content, covering fraction of Lyman Limit Systems (LLS: $N_{\rm HI} > 10^{17}$ cm$^{-2}$), and smaller \ion{H}{1}-bearing clouds.  Note that all projection and slice plots throughout this work use the same orientation and projection angle for consistency.}
\label{fig:HI_zoom}
\end{figure*}

\subsection{The \textsc{tempest} Simulations}
\label{subsec:tempest}

The \textsc{tempest} simulations are a set of cosmological hydrodynamic zoom simulations utilizing EHR to model the ctic medium of an L* galaxy.  For these simulations, we use the open-source hydrodynamics code \textsc{enzo} \citep{bryan:2014} along with modifications to implement EHR.  We follow the standard \textsc{enzo} treatment of cosmological zoom simulations described in more detail in \citet{hummels:2012} but briefly summarized here.  \textsc{enzo} represents hydrodynamics as a grid with variable resolution according to a density-based adaptive mesh refinement criterion coupled with EHR.  The equations of gas dynamics are solved using the piecewise parabolic method (PPM), a Godunov method that is third-order accurate in space \citep{colella:1984} including a nonlinear Riemann solver for better treatment of shocks.  \textsc{enzo} treats dark matter and stellar populations as collisionless particles modeled with an N-body adaptive particle-mesh gravity solver \citep{efstathiou:1985, couchman:1991}.  We include a prescription for star formation based on \citet{cen:1992} wherein dense, cool gas is periodically converted to stellar population particles when star forming conditions are met.  For the simulations presented in this paper, stellar feedback is parameterized as a thermal process, wherein young star particles return thermal energy, mass, and metals to the 3x3x3 grid of surrounding cells as the stellar population begins to form type II supernovae.

These simulations employ the \textsc{grackle} cooling libraries \citep{smith:2017} to account for the effects of metal cooling and to instantaneously follow all species of hydrogen and helium in non-equilibrium including collisional ionization, Compton cooling, recombination, brehmstrallung, photoionization, and photoexcitation.  For these calculations and to account for photoionization from the unresolved metagalactic component, we use the UV background spectrum from \citet{haardt:2012}.

The \textsc{tempest} initial conditions were derived from the \textsc{legacy} simulation project (Smith, O{\~n}orbe, \& Khochfar, in preparation).  The halo was selected from a dark-matter-only simulation with 2048$^3$ particles in a 100 Mpc/h box performed with a modified version of \textsc{gadget-2} \citep{springel:2001}.  It was selected as a Milky Way analog with no major mergers (10:1) after redshift 2, resulting in a galaxy with disky structure and virial mass of $\sim$$10^{12}$ M$_{\rm \odot}$ by $z=0$.  The halo was further chosen to reside in an \emph{average} galactic environment relative to the full simulation volume, based on distances to its 10 closest neighbor halos.

The \textsc{tempest} simulations presented in this paper follow the evolution of this single L* galaxy from $z=100$ to $z=0$ with analyses primarily at $z=1$.  We begin the simulation using only a density-based AMR scheme for managing the simulation resolution and run it forward to $z=3$.  At $z=3$, we restart the simulation with four variations: one continuing with standard density-based AMR, and three versions additionally including EHR at the different resolutions of 2.18, 1.09, and 0.545 comoving kpc (i.e., 2 kpc, 1 kpc, 500 pc used throughout this work).  These are run forward to $z=0$.  Each simulation including EHR places four nested ``must-refine'' boxes centered on the galaxy with widths of 200, 400, 800, and 1600 comoving kpc where each larger box drops the ``must-refine'' resolution by a factor of two (see Figure \ref{fig:refinement}).

\begin{figure*}[t]
\centering
\includegraphics[width=2.1\columnwidth, clip]{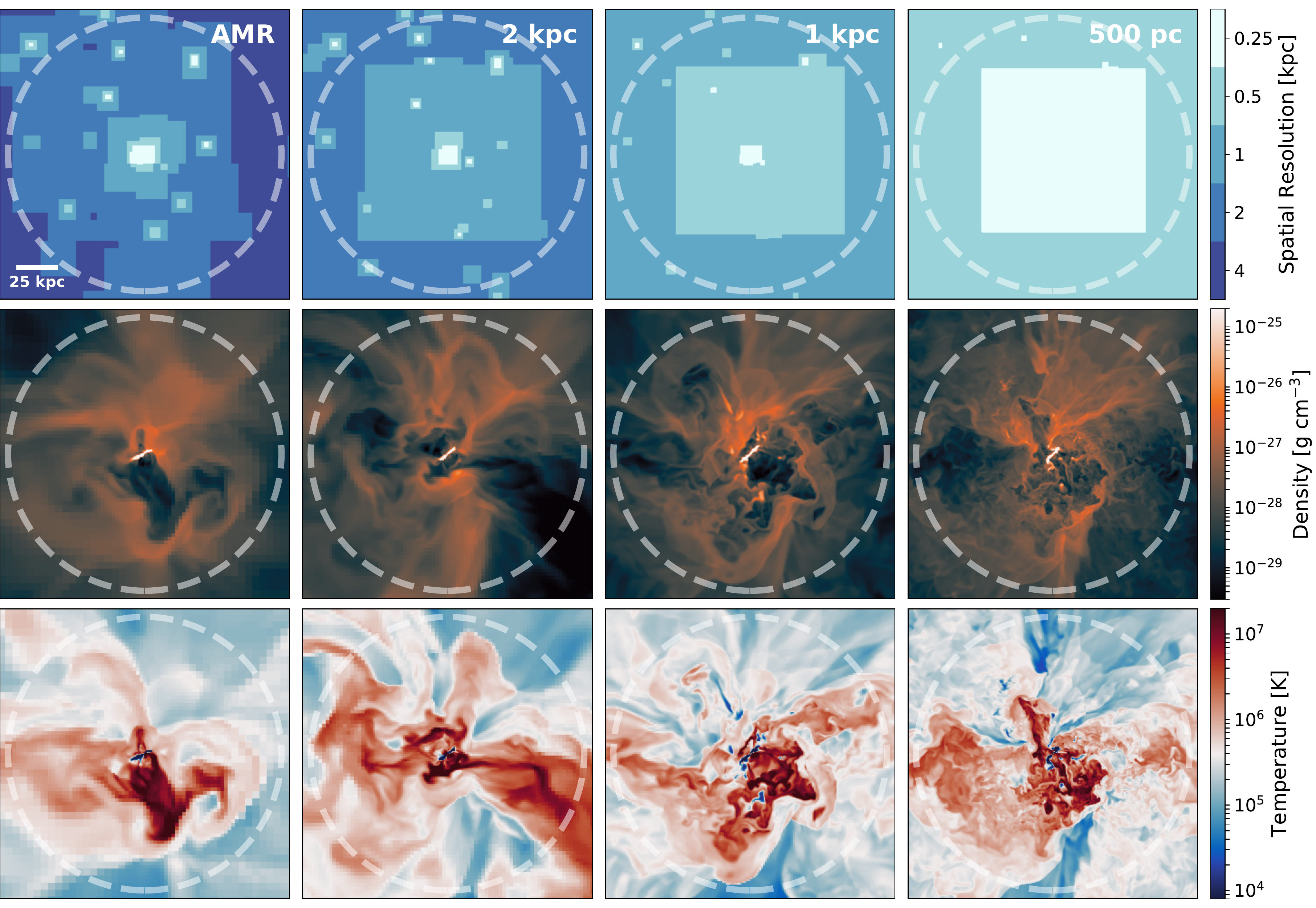}
\caption{Slices in physical quantities of a simulated L* galaxy at $z=1$ from the \textsc{tempest} simulations employing increasing levels of spatial resolution. \textbf{Top Row}: Effective spatial resolution maps with same orientation and colorbar as Figure \ref{fig:refinement}.  N.B., spatial resolution is in physical kpc at $z=1$, whereas runs are organized in comoving units (i.e., 2 comoving kpc = 1 physical kpc at $z=1$). \textbf{Middle Row}: Gas density. \textbf{Bottom Row}: Gas temperature. Images are 180 kpc on a side, pass through the center of the galaxy, and show $r_{\rm 200}$ as a dashed, white circle.  Enhancing the halo resolution results in a wider range of densities and temperatures with increased small-scale structure.}
\label{fig:slices}
\end{figure*}

\subsection{Calculating Ion Densities for \ion{H}{1} and \ion{O}{6}}

Throughout this work, we make comparisons between the \textsc{tempest} simulations and observations of \ion{H}{1} and \ion{O}{6}.  \textsc{enzo} follows the abundance for all species of hydrogen and helium (including \ion{H}{1}) in each hydrodynamic cell according to the \textsc{grackle} cooling libraries \citep{smith:2017} accounting for various sources of cooling and ionization described in Section \ref{subsec:tempest}.  

For all other ions, including \ion{O}{6}, we employ the \textsc{trident} code \citep{hummels:2017} to estimate their abundance in post-processing.  \textsc{trident} possesses a large lookup table assembled from hundreds of thousands of \textsc{cloudy} \citep{ferland:1998} radiation-hydrodynamics simulations to determine the inferred density of any ion based on a gas cell's density, temperature, metallicity, and redshift by accounting for collisional ionization and photoionization from a metagalactic background.  For the \textsc{tempest} simulations, we use the metagalactic background described in \citet{haardt:2012}.  See Figure \ref{fig:ion_fractions} for a visual representation of the ionization fraction for \ion{H}{1} and \ion{O}{6} as calculated by \textsc{trident}.

\section{Effects of Enhanced Halo Resolution}
\label{sec:discussion}

The impact that EHR has on the modeling of the CGM is significant and vividly apparent through visual inspection, as shown in Figure \ref{fig:HI_zoom}.  The simple increase in spatial resolution changes a number of properties of the CGM: it (1) more continuously and more correctly samples the various properties of the CGM (e.g., density, temperature, metallicity) leading to a broader range of these properties enabled by progressively smaller fluid elements; and (2) changes the thermal balance of the CGM, permitting more cool gas and less hot gas to exist.  The observational consequences of these physical changes mean that (3) EHR changes the ionic composition of the CGM, increasing low ion content while decreasing abundance of high ions; and (4) EHR decreases the size of individual ion-bearing clouds, enabling finer features in corresponding observations in images and spectra.

This paper focuses primarily on the changing thermal balance of the CGM and its observational repercussions (effects 2, 3 \& 4), while using effect 1 to explain \emph{why} this thermal balance changes.  The concurrent work by \citet{peeples:2018} explores the various effects of how smaller cloud sizes impact the spectral observables (4) associated with EHR.

It is also worth noting that the bulk galaxy properties are \emph{unaffected} by the change in resolution.  These quantities, including total mass, stellar mass, and gas mass, are included in Table \ref{tab:halo}.  We expect the slight changes in these values to be due mostly to the non-linear behavior of simulations due to stochastic sampling \citep{genel:2018}.

\begin{table}[h!]
\begin{center}
\begin{tabular}{ c c c  c  c  c} 
& AMR & \multicolumn{3}{c}{Enhanced Halo Resolution}\\
& Variable Res & 2.0 kpc & 1.0 kpc & 0.5 kpc \\
\hline
$m_{\rm 200}$ & $2.7 \times 10^{11}$ M$_{\rm \odot}$ & -1.1\% & +0.6\% & +0.3\% \\
$m_{\rm gas}$ & $1.6 \times 10^{10}$ M$_{\rm \odot}$ & -11\% & -2.3\% & -5.6\% \\
$m_{\rm stars}$ & $2.0 \times 10^{10}$ M$_{\rm \odot}$ & +0.1\% & +0.8\% & +3.1\% \\
$r_{\rm 200}$ & 84 kpc & -0.1\% & +0.4\% & +0.6\% \\
\end{tabular}
\caption{Masses of different components of the \textsc{tempest} galaxies at $z\sim1$. We used \textsc{hop} \citep{eisenstein:1998} and \textsc{yt} to calculate the halo properties median-filtered over ten consecutive simulation outputs centered on $z=1$. For the traditional AMR simulation, the galactic mass components are stated explicitly, while the EHR runs show their values relative to the AMR case to highlight any differences. EHR has little effect on bulk properties of the galaxy.}
\label{tab:halo}
\end{center}
\end{table}

\subsection{Small-Scale CGM Structure Changes}

Perhaps unsurprisingly, EHR leads to increased structure in the properties of the gas, and increasingly small coherent cloud sizes.  Figure \ref{fig:slices} illustrates this effect by comparing slices of the density and temperature fields centered on the galaxy at different resolutions.  As an additional reference, the top row shows the spatial resolution of the simulations in physical units at $z=1$.  At first glance, the bulk properties of the gas do not appear to change appreciably, but its small scale structure changes dramatically.  As spatial resolution increases, gas of different phases is allowed to exist in closer proximity, and more turbulent structure is apparent.  The extremes in the gas density and temperature increase.  This subtle point has significant ramifications to be discussed in Section \ref{sec:why}.  

\begin{figure*}[tb]
\centering
\includegraphics[width=2.1\columnwidth, clip]{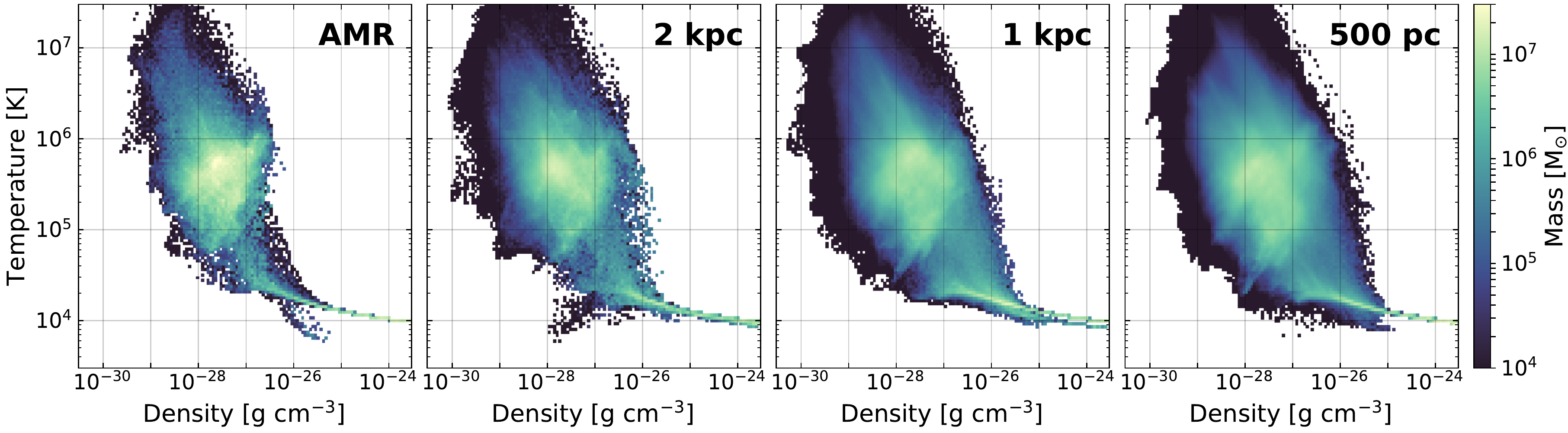}
\caption{Gas density-temperature phase diagrams of the simulated CGM across different resolutions.  Phase diagram includes gas in the hollow spherical shell from 10 kpc $< r < r_{\rm vir}$ for each simulation at $z=1$.  Increased resolution better samples the full range of gas density and temperature as seen in Figure \ref{fig:slices}.}
\label{fig:phase}
\end{figure*}

\begin{figure}[tb]
\centering
\includegraphics[width=1.0\columnwidth, clip]{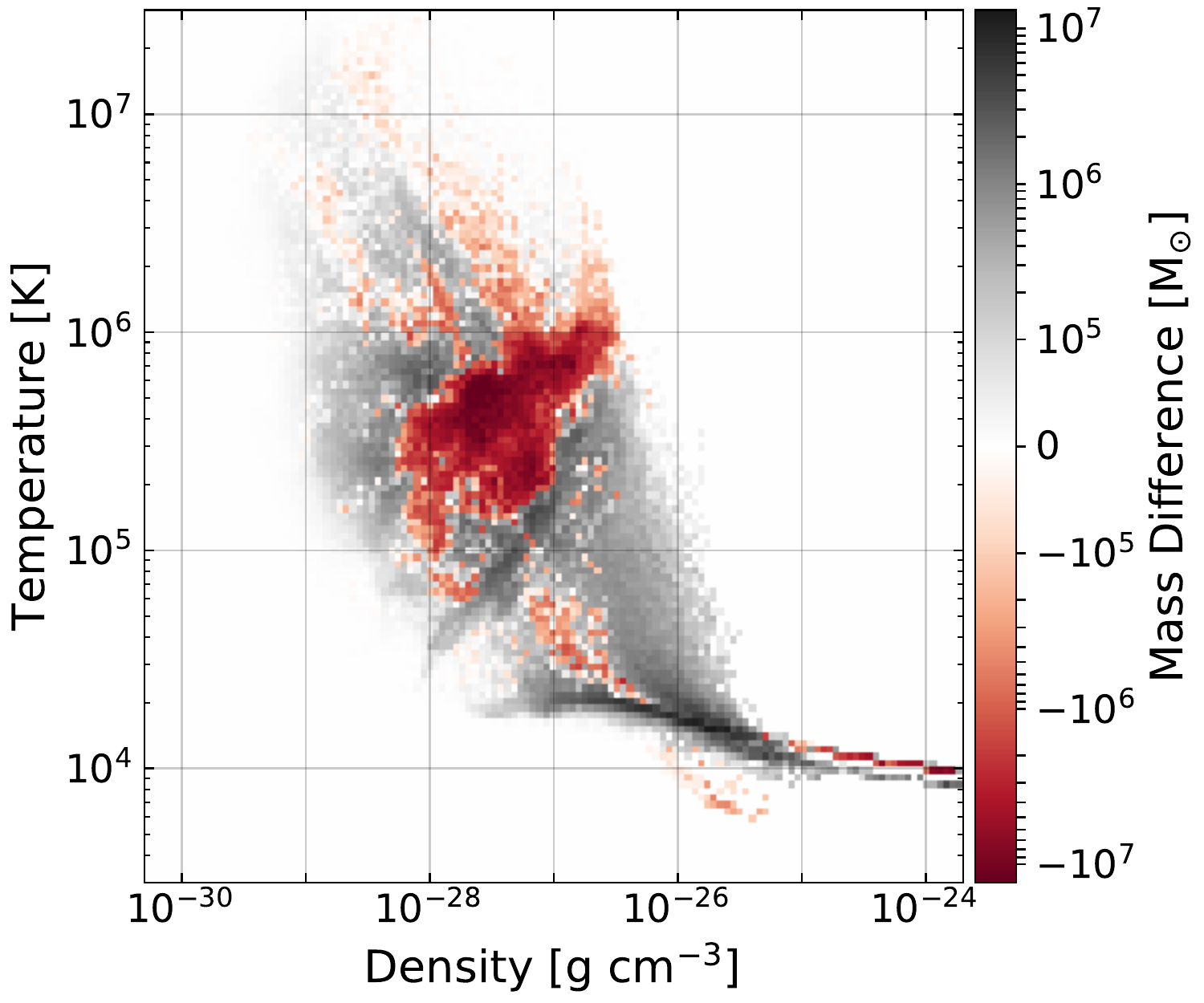}
\caption{Differences between the CGM gas density-temperature phase diagrams in Figure \ref{fig:phase} for the AMR simulation and the 1 kpc EHR simulation.  Gas in red is present in the AMR simulation and absent from the EHR simulation, whereas the reverse is true for gas in black.  Increased resolution effectively transforms gas from the red regions ($~5\times10^5$ K) to the black regions ($~2\times10^4$ K).  Explanation for what drives this effect is explored in Section \ref{sec:why}.}
\label{fig:phase_diff}
\end{figure}

\subsection{Changing Thermal Balance of the CGM}

To better assess how the gas density and temperature change with resolution, Figure \ref{fig:phase} shows a phase diagram plotting the CGM gas mass in bins of density and temperature.  Note that these plots only include gas in the spherical shell from 10 kpc to $r_{\rm vir}$ in each target galaxy.  (While the target galaxy's disk is excluded, we still observe some ISM gas from the satellites currently merging.)  Two effects can be noticed.  Firstly, the higher resolution simulations have a broader spread across the diagram, better quantifying how gas at higher resolution is better sampled in density and temperature, avoiding unnatural mixing of disparate phases imposed by coarse grid cells.  Secondly, there is a shift of material from the warm, low-density peak of the diagram to lower temperatures and higher densities.  This effect is more easily seen by taking a difference of two of these phase diagrams.  Figure \ref{fig:phase_diff} represents the difference between the AMR phase plot and the 1 kpc phase plot from Figure \ref{fig:phase}.  Interestingly, the warm, low-density gas highlighted in red is present in the low resolution simulation but absent from the high-resolution simulation.  Conversely, the cool, dense material shown in gray/black is present in the high-resolution simulation but absent from the low-resolution simulation. Understanding this behavior will occupy much of the remainder of this study.

Table \ref{tab:CGM} further quantifies how much resolution changes the thermal balance of the CGM by calculating the mass of each phase of gas in the CGM (10 kpc $< r < r_{\rm vir}$) of snapshots at $z=1$.  Cool gas ($10^4 < T < 10^5$ K) content increases with resolution, whereas warm gas ($10^5 < T < 10^6$ K) has a slight decrease with resolution.  These trends are somewhat noisy due to our small sample size and stochastic effects.

\begin{table}[ht!]
\begin{center}
\begin{tabular}{ c c c  c  c} 
& AMR & \multicolumn{3}{c}{Enhanced Halo Resolution}\\
& Variable Res & 2.0 kpc & 1.0 kpc & 0.5 kpc \\
\hline
$m_{\rm CGM}$ & $8.0 \times 10^{9}$ M$_{\rm \odot}$ & -13\% & -13\% & -4\% \\
$m_{\rm cold}$ & $4.4 \times 10^{8}$ M$_{\rm \odot}$ & +52\% & +12\% & +83\% \\
$m_{\rm cool}$ & $7.5 \times 10^{8}$ M$_{\rm \odot}$ & +37\% & +108\% & +84\% \\
$m_{\rm warm}$ & $6.2 \times 10^{9}$ M$_{\rm \odot}$ & -24\% & -29\% & -21\% \\
$m_{\rm hot}$  & $5.2 \times 10^{8}$ M$_{\rm \odot}$ & +7\% & -13\% & +5\% \\
$m_{\rm HI}$ & $2.4 \times 10^{8}$ M$_{\rm \odot}$ & +40\% & +7\% & +59\% \\
$m_{\rm OVI}$ & $4.0 \times 10^{4}$ M$_{\rm \odot}$ & -12\% & -33\% & -22\% \\
\end{tabular}
\caption{Masses of different components of the CGM of the \textsc{tempest} galaxies at $z\sim1$. All quantities are measured from 10 kpc $< r < r_{\rm vir}$ to remove the core galaxy.  Components are median sampled over 10 consecutive simulation outputs centered on $z=1$.  For the AMR simulation, mass of each component is listed, whereas for the EHR simulations we denote a percentile describing how each component changes relative to the AMR simulation.  Different temperature components are defined as $T_{\rm cold} \in (0, 10^4]$K, $T_{\rm cool} \in [10^4, 10^5]$K, $T_{\rm warm} \in [10^5, 10^6]$K, $T_{\rm hot} \in [10^6, \infty)$K.  EHR changes the thermal balance of the CGM, favoring cool gas over warm gas and low ions (e.g., \ion{H}{1}) over high ions (e.g., \ion{O}{6}).}
\label{tab:CGM}
\end{center}
\end{table}

There appears to be some mechanism transforming the thermal balance of the CGM, enabled by higher spatial resolutions.  However, from these diagnostics alone, it is unclear if resolution enables gas to change from warm to cool gas, or rather if resolution prevents existing cool gas from being spuriously transformed into warm gas.  \emph{Why} EHR changes the gas is further discussed in Section \ref{sec:why}.

\begin{figure*}
\centering
\includegraphics[width=2.1\columnwidth, clip]{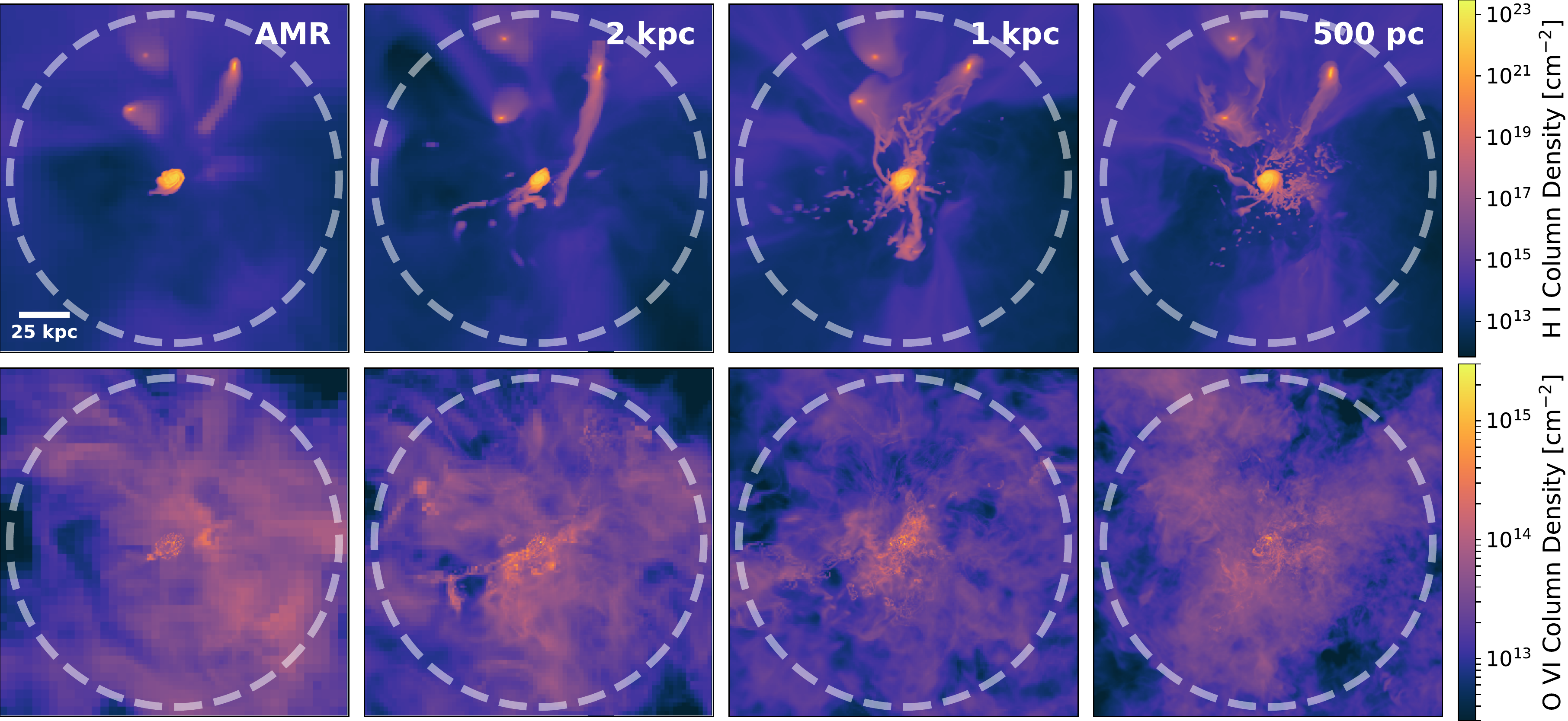}
\caption{Column density projections of a simulated L* galaxy at $z=1$ employing different levels of spatial resolution. \textbf{Top Row}: Neutral hydrogen column density maps. \textbf{Bottom Row}: \ion{O}{6} column density maps.  Enhancing the halo resolution contributes to additional neutral hydrogen content throughout the halo. This is most evident as an increase in Lyman Limit Systems ($N_{\rm HI} > 10^{17}$ cm$^{-2}$) found in progressively smaller clouds in the inner halo, the shredded remnants of merging galaxies, supernova-driven outflows, and cosmological filaments.  Increases in \ion{H}{1} content are accompanied by a nearly imperceptible decrease in \ion{O}{6} global column densities.}
\label{fig:projections}
\end{figure*}

\subsection{Increased \ion{H}{1} Content and Decreased \ion{O}{6} Content}

The change in the thermal state of the halo gas enabled by EHR directly impacts the observational characteristics of the CGM.  The CGM is detected through emission and absorption of the ionic species that compose it.  Different ions reside in different density and temperature regimes of the CGM gas according to the energy needed to ionize each species.  Figure \ref{fig:ion_fractions} shows the ion fraction plots for \ion{H}{1} and \ion{O}{6}, indicating the phases of gas where these ions most frequently occur plotted on the same axes as Figures \ref{fig:phase} and \ref{fig:phase_diff}.  As noted, \ion{H}{1} and other low ions reside in cool, dense gas, whereas \ion{O}{6} and other high ions typically exist in higher temperature gas.

Because of the additional cool, dense gas present in the higher resolution simulations, we expect to see an increase in the low ion content of the gas and a deficit in the high ion content of the gas.  Figure \ref{fig:projections} displays column density maps in \ion{H}{1} and \ion{O}{6} for the \textsc{tempest} galaxies of different spatial resolution.  This figure illustrates how the \ion{H}{1} abundance increases in regions of higher spatial resolution while it decreases the \ion{O}{6} content to a lesser extent.  Aside from overall abundance, we see distinct changes in the observational gas properties.  The \ion{H}{1} resides in increasingly narrow structures, primarily in the inner halo, generally associated with disrupted merging galaxies.  \ion{O}{6} on the other hand is much more well distributed throughout the halo with larger cloud sizes.

\begin{figure*}
\centering
\includegraphics[width=2.1\columnwidth, clip]{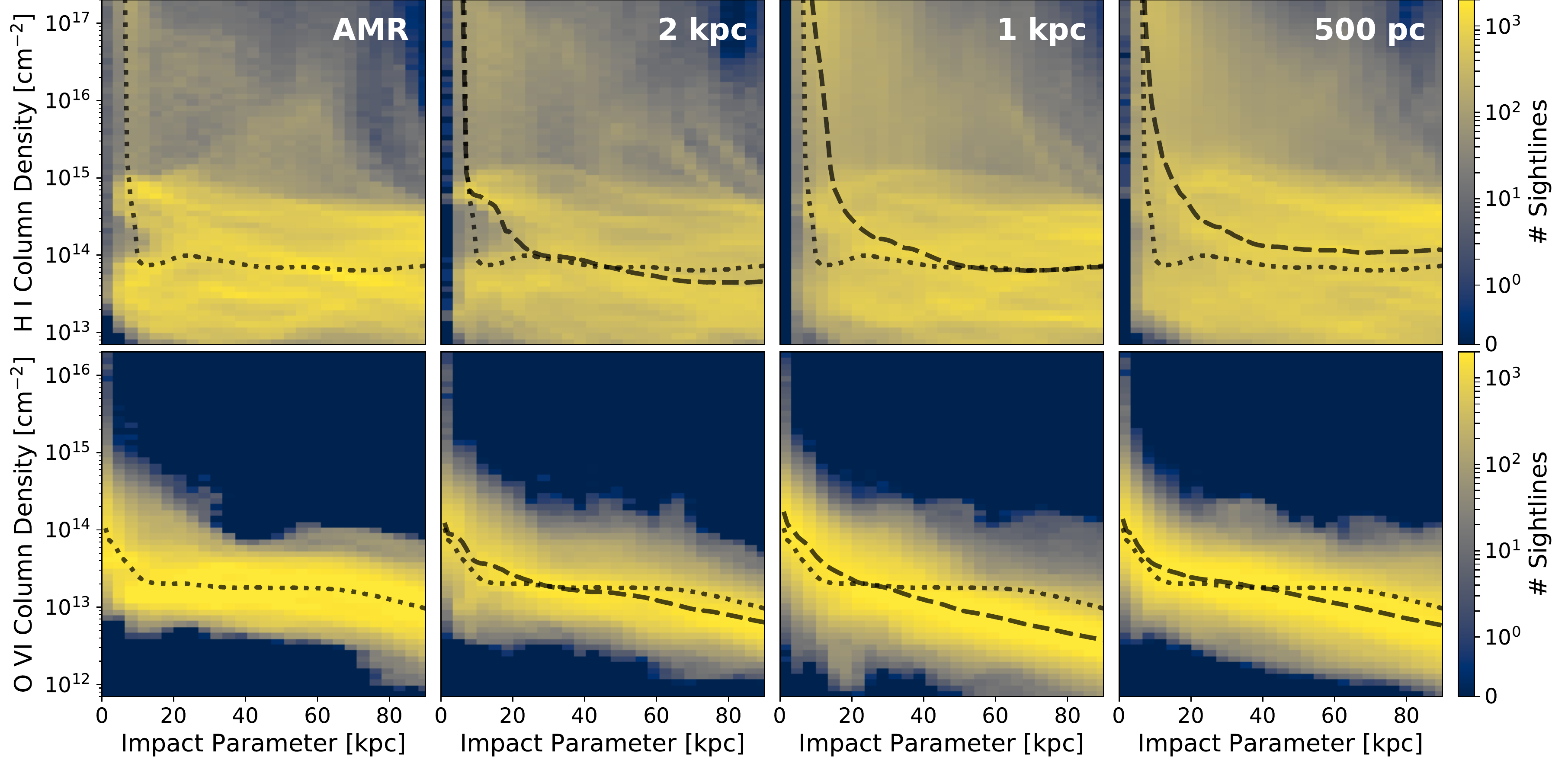}
\caption{Radial profiles of \ion{H}{1} (\textbf{top}) and \ion{O}{6} (\textbf{bottom}) column density versus impact parameter.  While the heatmap represents the full distribution of sightlines, the black lines indicate the median column density value at each impact parameter (dotted = AMR, dashed = EHR).  Profiles were generated for each pixel in column density projections like Figure \ref{fig:projections}, where each sightline is a line integral of ion number density.  We used consecutive simulation outputs and multiple projection angles to avoid spatial and temporal biases in our profiles.  Increasing resolution correlates with higher \ion{H}{1} columns densities and a more moderate decrease in \ion{O}{6} column densities.}
\label{fig:profiles}
\end{figure*}

Figure \ref{fig:profiles} further quantifies the increased \ion{H}{1} and slightly decreased \ion{O}{6} abundance when halo gas attains higher spatial resolution.  Figure \ref{fig:profiles} consists of radial profiles plotting column density as a function of projected distance from the galactic center (i.e., impact parameter).  The radial profiles were produced from column density projections like those in Figure \ref{fig:projections}, sampling multiple projection angles over twenty simulation outputs spanning a gigayear around $z=1$ to wash out any spatial or temporal biases.  The heat map shows the full distribution of column densities from every sightline (i.e., pixel) in the sampled images with a black line representing the median column density value at each impact parameter.  

In the top row, we see that increasing resolution leads to an increase in high-column density \ion{H}{1} systems e.g., Lyman Limit Systems ($N_{\rm HI} > 10^{17}$ cm$^{-2}$), which raises the median column density value, primarily in the interior 40 kpc (about half the virial radius).  However, the entire median \ion{H}{1} column density profile is boosted at all impact parameters for our highest-resolution (500 pc) simulation, increasing observed \ion{H}{1} column densities by a factor of two throughout the halo.  No convergence in this behavior is seen, suggesting that additional EHR will lead to even larger observed \ion{H}{1} column densities.

The effects of resolution on \ion{H}{1} are in contrast to those effects on \ion{O}{6}.  Increasing resolution seems to slightly \emph{decrease} the column densities of \ion{O}{6}, although the effect seems less significant than for \ion{H}{1} and primarily in the outer part of the galactic halo.  

While we do not show them here, the behavior of \ion{H}{1} is representative of all of the low ions (e.g., \ion{Mg}{2}, \ion{C}{2}, \ion{N}{2}), and \ion{O}{6} is representative of all the high ions in terms of their behavior with increases in resolution.  This fact is due to the similarity of ionization fraction functions (Figure \ref{fig:ion_fractions}) between species having similar ionization potentials.

\subsection{Progressively Smaller \ion{H}{1} Cloud Sizes}
\label{subsec:size}

In addition to an increase in total \ion{H}{1} content and observed \ion{H}{1} column density, EHR leads to a decrease in the size of the \ion{H}{1}-bearing clouds.  This can be seen by eye in Figures \ref{fig:HI_zoom} and \ref{fig:projections}, but we quantify it more precisely using a clump-finding algorithm.  We employ the clump finder implemented in \textsc{yt} \citep{turk:2011} and fully described in \citet{smith:2009}, giving a brief overview of the method here.  

Much like many halo finders operate on the spatially-varying gravitational potential, the clump finder uses a contouring algorithm to identify unique, topologically-connected structures in gas density.  In our usage, it first creates a single contour at the $n_{\rm HI} > 10^{-8}$ cm$^{-3}$ threshold, the lowest density we found necessary to identify partial Lyman Limit Systems and Lyman Limit Systems (pLLS: $N_{\rm HI} > 10^{16}$ cm$^{-2}$), identifying spatially contiguous clouds that are above this density threshold.  In subsequent steps, this threshold is doubled until we reach the maximum \ion{H}{1} density of $n_{\rm HI} > 10^{2}$ cm$^{-3}$.  Isolated structures are identified as separate contours through recursion.  We filter out clumps with fewer than 10 cells or $m_{\rm HI} < 1$ M$_{\odot}$ to avoid noise.  The result is a collection of unique \ion{H}{1} clouds identifying all the pLLS and LLS structures throughout the halo.  

The clump finder was run on the $z=1$ snapshot for each simulated galaxy, identifying \ion{H}{1} clumps from 10 kpc $< r < r_{\rm vir}$.  Each clump was cataloged by its \ion{H}{1} mass $m_{\rm HI}$, its shortest spatial dimension $l_{\rm short}$, and its longest spatial dimension $l_{\rm long}$.  The median values for these quantities in the different \textsc{tempest} simulations are presented in Table \ref{tab:clumps}.  

\begin{table}[h!]
\begin{center}
\begin{tabular}{ c c c c  c  c  c} 
& & AMR & \multicolumn{3}{c}{Enhanced Halo Resolution}\\
& & Variable & 2.0 kpc & 1.0 kpc & 0.5 kpc \\
\hline
$n_{\rm clouds}$ & & 7 & 25 & 137 & 443 \\
$l_{\rm short, med}$ & [kpc] & 4.4 & 4.4 & 2.2 & 1.4 \\
$l_{\rm long, med}$ & [kpc] & 13 & 7.7 & 4.4 & 2.5 \\
$m_{\rm HI, med}$ & [$10^{2}$ M$_{\rm \odot}$] & 43 & 11 & 10 & 1.4 \\
\end{tabular}
\caption{Statistics on \ion{H}{1} clouds in the CGM of each \textsc{tempest} simulation. Quantities were determined using a clump finder on $n_{\rm HI}$ in the region 10 kpc $< r < r_{\rm vir}$ to identify contiguous structures with $10^{-8}$ cm$^{-3} < n_{\rm HI} < 10^{2}$ cm$^{-3}$.  Cloud quantities include median size of cloud in its short ($l_{\rm short, med}$) and long ($l_{\rm long, med}$) dimensions and \ion{H}{1} mass ($m_{\rm HI}$).  Individual \ion{H}{1} cloud mass and size decrease with increased resolution, while overall number of clouds increases.}
\label{tab:clumps}
\end{center}
\end{table}

Consistent with a visual inspection of Figure \ref{fig:projections}, Table \ref{tab:clumps} reveals that increased spatial resolution leads to a larger number of progressively smaller and lower-mass \ion{H}{1}-bearing clouds.  At each factor-of-two increase in spatial resolution, we find a roughly factor-of-two drop in our median cloud size in both its long and short dimensions.  Recall that our quoted resolutions are in co-moving units, so the actual physical scales that they achieve at this redshift $z=1$ are a factor of $1+z=2$ smaller.  This implies that the median \ion{H}{1} cloud in each simulation has 4-6 cells in its shortest dimension and 8-10 cells in its longest dimension, near the limit of what could be considered a resolved structure.  

Increases in resolution allow gas to exist in smaller structures.  These results confirm that our \ion{H}{1} clouds are fragmenting to smaller scales when possible.  Our findings are consistent with the theoretical work of \citet{mccourt:2018} on thermally unstable gas ``shattering'' to cool clouds on ever-smaller scales only converging at the sub-parsec level.  Like the increase in \ion{H}{1} column densities with spatial resolution, we do not find any convergence in the minimum size scale of \ion{H}{1}-bearing clouds.  We predict similar behavior for other low-ion-bearing clouds like \ion{Mg}{2}. However, the sizes of high-ion-bearing clouds like \ion{O}{6} are likely unaffected by EHR, since high ions tend to sit in hotter, low-density gas more stable against thermal instability and collapse.

\section{Why Does Enhanced Halo Resolution Work?}
\label{sec:why}

It is important to investigate \emph{why} increased resolution has the effects that it does on the evolution of the CGM, in part because it can reveal insight into other ways to better model the CGM, and how the resolution effects may converge.  We identify two separate mechanisms by which Enhanced Halo Resolution enables more cool, dense \ion{H}{1}-bearing gas to exist in the simulated CGM, as clearly demonstrated in the gas density-temperature phase diagrams of Figures \ref{fig:phase} and \ref{fig:phase_diff}.  The first mechanism employs increased spatial resolution to effectively transfer warm gas to a cooler, denser phase through cooling and condensation, whereas the second mechanism prevents already cool gas from being artificially heated.

\subsection{Mechanism A:\\EHR Better Samples Gas Properties \& Seeds Precipitation}
\label{subsec:mechanism_a}

The most obvious explanation for the increase in cool, dense gas and decrease in warm, low-density gas found in EHR simulations is that resolution somehow triggers cooling.  There is substantial theoretical work on the topic of efficient cooling of thermally unstable gas found in halos of clusters and galaxies, collectively described as ``precipitation'' \citep[e.g.,][]{mccourt:2012, sharma:2012, voit:2015}.  Generally, precipitation of cool gas out of the hot halo medium is predicted to occur when cooling of the gas becomes efficient relative to its local freefall time: $t_{\rm cool} / t_{\rm ff} \le 10$.  While the gas freefall time is dictated by the galactic potential, its cooling time is a function of the local properties of the gas (e.g., density, temperature, and metallicity), so substantial decreases in the cooling time of the gas should yield efficient cooling and precipitation.  $t_{\rm cool}$ decreases with increases in density and metallicity, whereas $t_{\rm cool}$ is a non-linear function of temperature, found to be lowest in the temperature range $2\times10^4$ K $< T < 10^6$ K, precisely where we see the effects of EHR in Figure \ref{fig:phase_diff}.

\begin{figure}[t]
\centering
\includegraphics[width=1.0\columnwidth, clip]{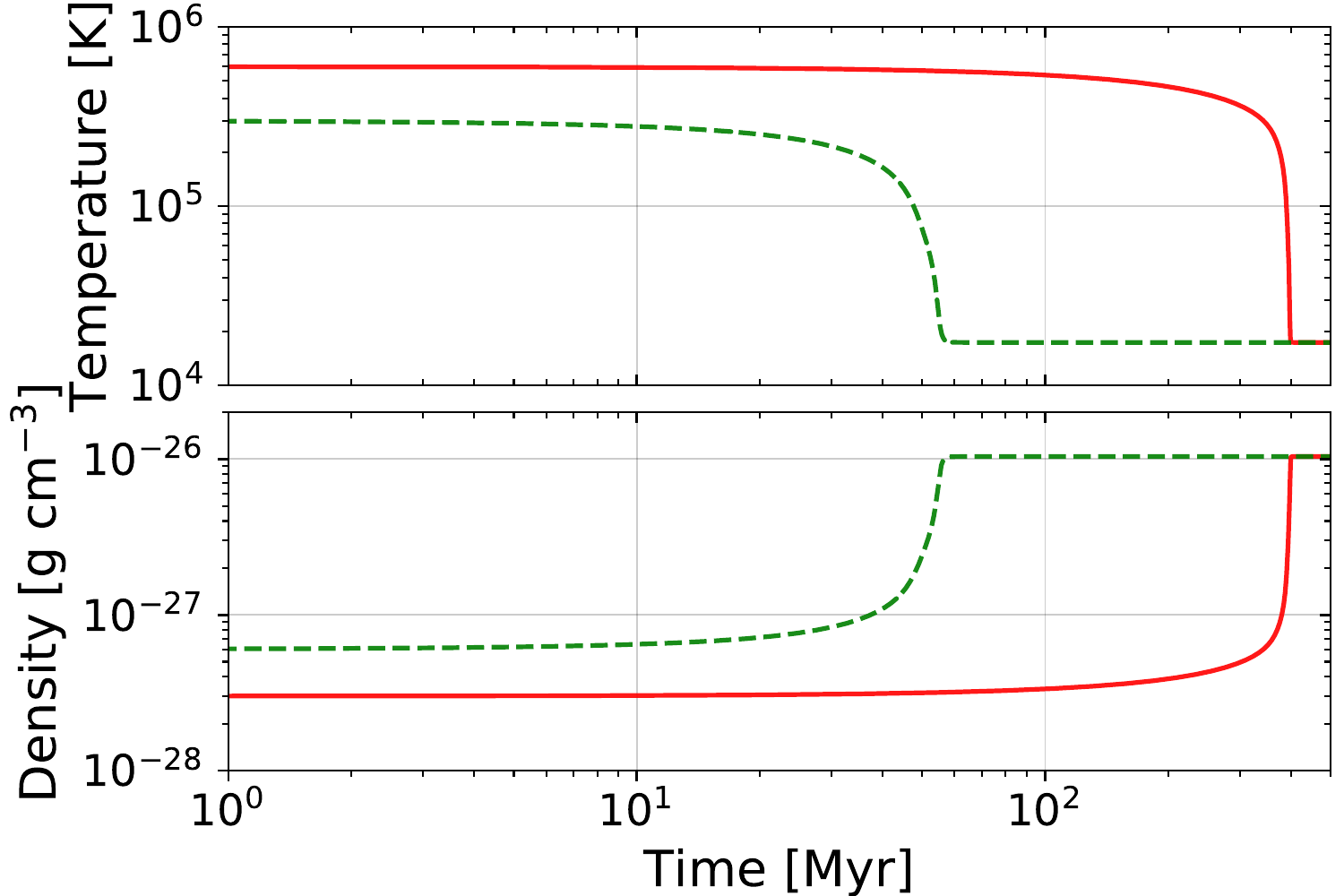}
\caption{Simple single-zone cooling simulations run with the \textsc{grackle} cooling code \citep{smith:2017}.  These two isobaric cooling models were run forward in time with metallicity = $0.05{\rm Z}_{\odot}$. The solid, red line represents the density, temperature, and metallicity of gas found near the virial radius of the \textsc{tempest} galaxies, indicating a cooling time of 400 Myr.  The dashed, green model doubles the initial density and correspondingly drops the temperature by two, which decreases the cooling time by a factor of $\sim$8.  These models demonstrate how sensitive cooling time is to initial gas properties.}
\label{fig:model}
\end{figure}

To demonstrate the sensitivity of gas cooling to its initial conditions, we ran two simple single-cell cooling models using the \textsc{grackle} code \citep{smith:2017}, shown in Figure \ref{fig:model}.  Each model follows the time evolution of a cloud of gas kept at constant pressure and allowed to cool.  The solid, red line shows the behavior of gas with properties representative of the galactic halo of the \textsc{tempest} galaxy at $z=1$: $\rho = 3 \times 10^{-28}$ cm$^{-3}$, $T = 6 \times 10^5$ K,  $Z=0.05 {\rm Z}_{\odot}$.  These gas properties also occupy the heart of the red region in Figure \ref{fig:phase_diff}, the phase of gas abundant in our low-resolution simulations but scarce in our high-resolution simulations.  The dashed, green line represents gas with a factor of two higher density and a corresponding factor of two lower temperature due to our isobaric condition.  The simple change of doubling the density has a dramatic impact on the cooling time of this gas, allowing it to cool about seven times faster in 55 Myr (green) compared to the baseline 400 Myr (red).  In both cases, the final density and temperature of the gas is $10^{-26}$ cm$^{-3}$ and $\sim2 \times 10^{4}$ K, exactly where the buildup of material occurs in the high-resolution simulations, shown in Figure \ref{fig:phase_diff} as black-gray.

Recall that EHR leads to increased sampling of various properties of the CGM gas, including its density, temperature, and metallicity.  In traditional AMR simulations lacking resolution in the halo, a parcel of gas that in reality exists in multiple phases is instead represented by a coarse grid cell averaging out over these multiphase conditions.  EHR minimizes this problem since the gas is more continuously and correctly sampled, resulting in a broader spread in the gas properties.  As demonstrated in our sample cooling tests in Figure \ref{fig:model}, even slight differences in gas phase can produce a non-linear impact on the gas cooling, potentially leading to a cooling runaway manifested as precipitation.

\begin{figure}[t]
\centering
\includegraphics[width=1.0\columnwidth, clip]{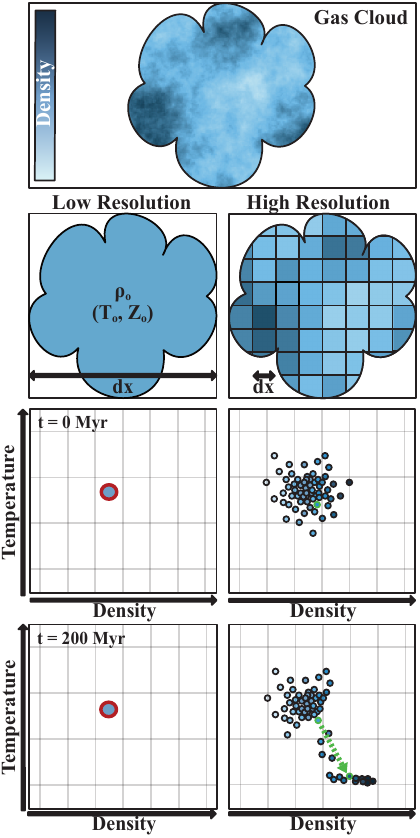}
\caption{Schematic of how EHR seeds cool gas precipitation.  \textbf{Top Panel}: Representation of a real gas cloud consisting of a continuous range of gas properties (e.g., density, temperature, metallicity). \textbf{Second Row}: Low- and high-resolution sampling of this gas cloud smears out the extremes in its gas properties by averaging them over resolution elements (as demonstrated in Figure \ref{fig:slices}). \textbf{Third Row}: These resolution elements are plotted in a phase diagram similar to Figure \ref{fig:phase}.  The low-resolution simulation only probes a single, average density and temperature value, whereas the high-resolution simulation samples a distribution of many values.  Red and green circles represent the red and green lines from our cooling models in Figure \ref{fig:model}. \textbf{Bottom Row}: After 200 Myr, the red circle shows no appreciable cooling, whereas
the slightly denser, cooler material in the high-resolution simulation efficiently cools and precipitates following the green dashed line to form a pile up of dense, cool material as in Figure \ref{fig:phase_diff}.}
\label{fig:schematic_a}
\end{figure}

\begin{figure}[t]
\centering
\includegraphics[width=1.0\columnwidth, clip]{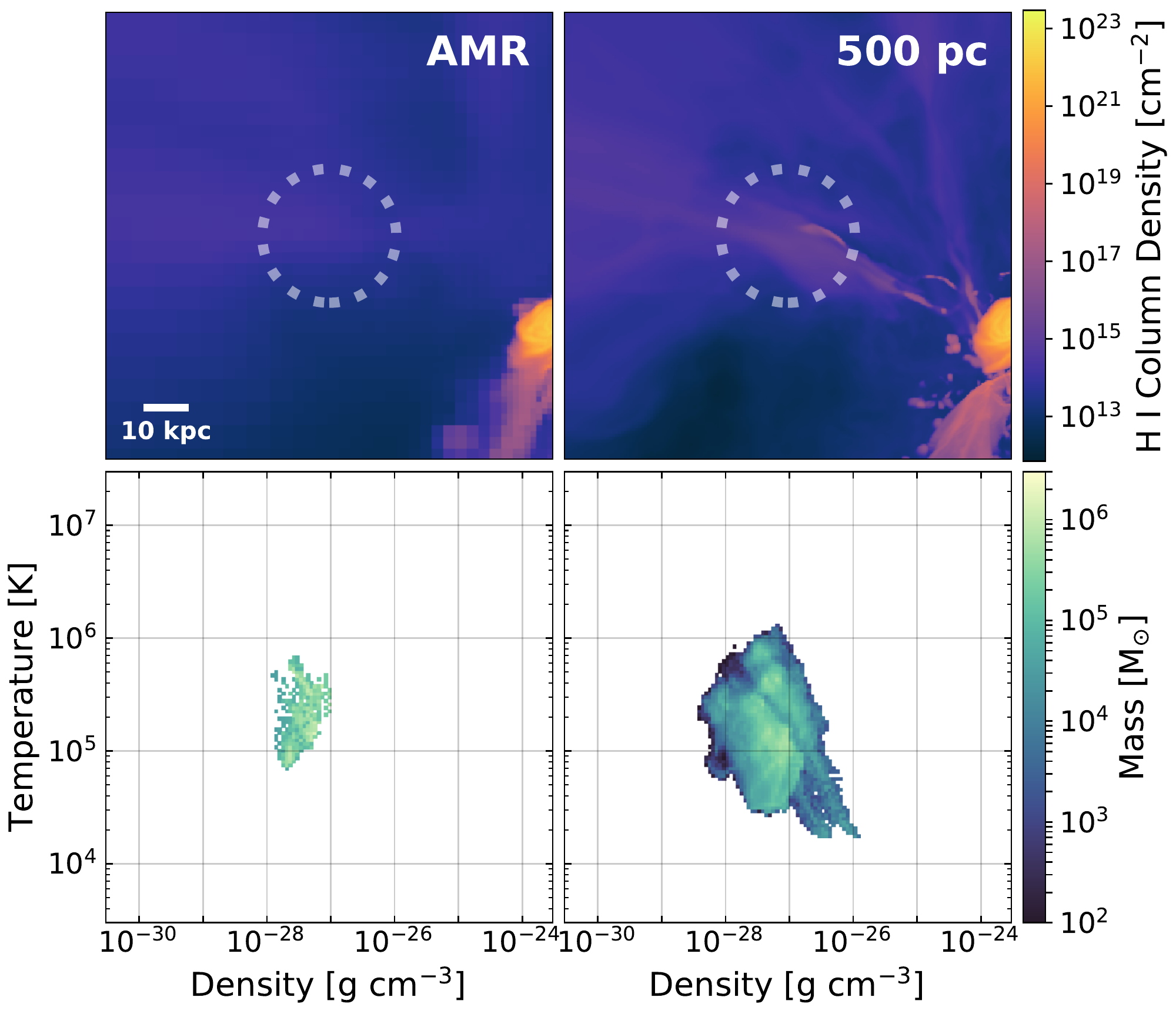}
\caption{Example of Mechanism A in action as cool gas condenses out of an inflowing filament in EHR simulation but not in traditional AMR simulation.  \textbf{Top Row}: \ion{H}{1} column density projections of \textsc{tempest} simulations at $z=0.7$ centered on an inflowing cool filament 55 kpc away from host galaxy (seen in lower right).  Dotted white circle represents a 30-kpc-wide sphere sampling the filament's gas for the phase diagram on the \textbf{Bottom Row}.  The finger-like LLS in the center-right part of the dotted white circle spontaneously cooled out of the inflowing filament.  The phase diagram reveals that this cloud is an order of magnitude cooler and denser than the corresponding gas in the AMR simulation, following the isobaric cooling path in Figure \ref{fig:schematic_a}.  Mechanism A is active in the EHR \textsc{tempest} simulations, primarily observed in stable, cool, dense structures like inflowing filaments.  A movie of the time-evolution of this plot can be found online$^{1}$.}
\label{fig:filament}
\end{figure}

Thus, we propose that the increased resolution of EHR better samples the intrinsic CGM gas quantities, and the resulting tail of the gas distribution with shorter cooling time cools non-linearly, culminating in precipitation of dense, cool gas in the halo. 
Figure \ref{fig:schematic_a} illustrates the effects of this proposed mechanism.  The top panel represents a ``real'' gas cloud consisting of a range of densities, temperature, and metallicities.  We then depict the low- and high-resolution versions of this cloud graphically (second row) and in density-temperature phase space (third row).  On the left, the unresolved cloud is the size of a single grid cell, so it only has a single density value, whereas on the right, the cloud has an 8x increase in resolution (typical of the differences between AMR and EHR).  Each phase diagram includes appropriately-placed red and green cells to represent the starting conditions of the two grackle cooling models illustrated as red and green lines in Figure \ref{fig:model}.  Consistent with the cooling models, after 200 Myr the high-density, low-temperature gas (green circle) of the higher-resolution cloud has radiated away its energy and settled into dense $\sim$10$^4$K gas in the lower right of the phase plot, whereas the poorly-resolved cloud (red circle) has not cooled appreciably.  This mechanism correctly reproduces the trend in Figure \ref{fig:phase_diff} by transferring gas from the warm, low-density regime to the cool, high-density regime through resolution-seeded precipitation.

We further confirm that the proposed mechanism of resolution-seeded precipitation is operating in the \textsc{tempest} simulations.  We identify a cosmological filament and observe it over time feeding the target galaxy with cool gas from the intergalactic medium.  At certain points during its evolution, the filament simulated with EHR undergoes transient precipitation of cool, dense clouds whereas no such activity occurs in the traditional AMR simulated filament.  This is illustrated in Figure \ref{fig:filament} at $z=0.7$.  The top panels show column density projections of \ion{H}{1} centered on the filament, 55 kpc distant from the primary galaxy seen in the lower right.  A white circle represents a 30-kpc-wide sphere sampling the gas properties of the filament, plotted as a density-temperature phase diagram in the bottom panels.  The narrow $\sim$1 kpc-wide finger of high density \ion{H}{1}-bearing gas found in the center right of the white circle in the EHR simulation is the precipitating cool gas.  

The phase diagram reveals that this precipitate is at least an order of magnitude denser and cooler in the EHR simulation than it is in the AMR simulation.  It is notable how the phase diagrams resembles those in the bottom of schematic Figure \ref{fig:schematic_a}, with gas following isobaric lines from top left to bottom right, revealing the smoking gun of runaway cooling.  Readers are encouraged to watch the full time evolution of this filament as a movie\footnote{Full filament movie available at: \href{http://chummels.org/tempest}{http://chummels.org/tempest}}.

\begin{figure}[t]
\centering
\includegraphics[width=1.0\columnwidth, clip]{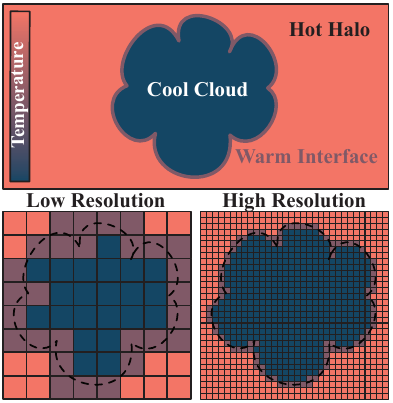}
\caption{Schematic of how EHR prevents cool gas from evaporating.  \textbf{Top Panel}: Cartoon of cool gas residing in a hot halo with a thin warm interface between the two.   \textbf{Bottom Panels}: Low- and high-resolution grids approximate this cloud with mixing along its boundary proportional to the grid scale.  Poorly-resolved clouds artificially mix with the surrounding medium on shorter timescales, whereas well-resolved cool clouds are preserved longer as they would be in reality. The result is more cool gas in EHR simulations correctly reproducing the behavior of the \textsc{tempest} simulations in Figure \ref{fig:phase_diff}.}
\label{fig:schematic_b}
\end{figure}

\subsection{Mechanism B:\\EHR Prevents Artificial Mixing of Cool Gas}
\label{subsec:mechanism_b}

Computational hydrodynamics simulations work by discretizing the continuous spatial distribution of fluid into distinct resolution elements.  This process functions well when the size of the resolution elements is small relative to the natural size of fluid structures, but when fluid structures are poorly resolved, numerical artifacts arise.  In general, Eulerian (i.e., grid-based) codes like the one used for the \textsc{tempest} simulations tend to be overly diffusive on small resolution scales, washing out structures near the resolution limit through excessive mixing with surrounding fluid elements.  In contrast, Lagrangian (i.e., particle-based) codes generally under-predict the amount of fluid mixing occurring on small scales, locking material into artificially segregated clouds at the resolution scale \citep[e.g.][]{agertz:2007}.  Either way, unphysical effects begin to occur at scales near the resolution limit of the simulation.

\begin{figure}
\centering
\includegraphics[width=1.0\columnwidth, clip]{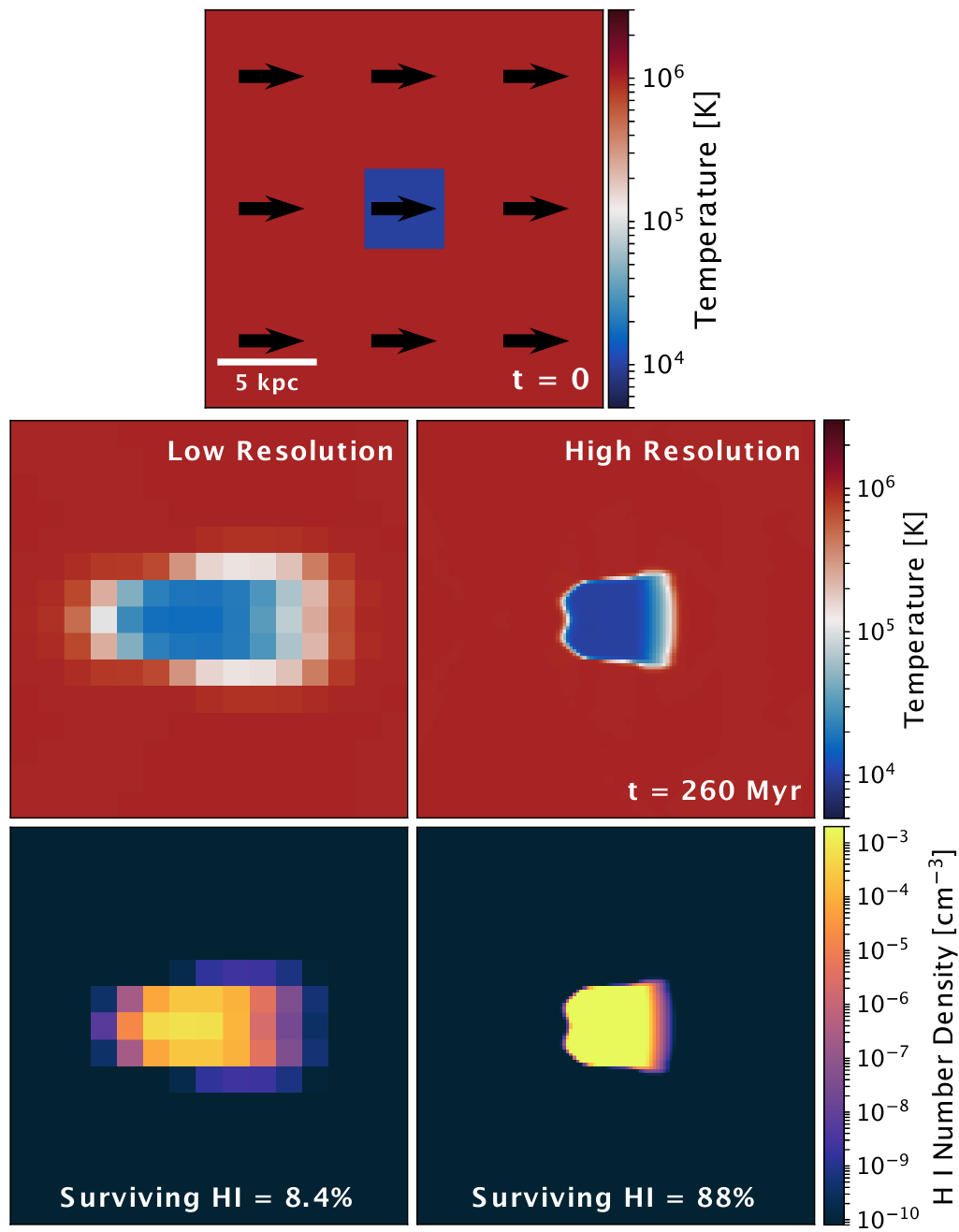}
\caption{Hydrostatic square advection test demonstrating the diffusive nature of underresolved cool clouds and their rapid depletion of \ion{H}{1}.  \textbf{Top Panel}: Our initial conditions place a 4-kpc-wide cool (T = 10$^4$K), dense ($\rho$ = 10$^{-26}$ g cm$^{-3}$) cloud in pressure equilibrium with a hot (T = 10$^6$K), low-density ($\rho$ = 10$^{-28}$ g cm$^{-3}$) medium, with all fluid moving at constant velocity ($v_{\rm x}$ = 150 km/s) similar to cool filament inflow velocities in \textsc{tempest} simulations. After one domain-crossing time, 260 Myr, we show (\textbf{Middle Row}) temperature and (\textbf{Bottom Row}) \ion{H}{1} density slices for a (\textbf{Left}) low-resolution simulation initially resolved by 3 1.33 kpc cells, and (\textbf{Right}) high-resolution simulation with 8x higher spatial resolution.  The low-resolution cool clouds mix quickly with their surroundings through numerical diffusion, which helps to explain the behavior of EHR preserving cool gas and \ion{H}{1} content in the halo similar to what is shown in our schematic Figure \ref{fig:schematic_b}.}
\label{fig:blob}
\end{figure}

Figure \ref{fig:schematic_b} illustrates this effect for a grid-based representation of a cool cloud entrained in hot halo gas.  The top panel represents how, in reality, a cool cloud slowly evaporates through its warm interface with the surrounding hot medium due to thermal conduction processes (conduction is excluded from our simulations for simplicity).  The bottom row shows the grid-based representation of this scenario for both low- and high-resolution cases.  The grid cells bordering along the edge of the cloud contain both cool and hot gas, mixing to form a warm interface.  In the low resolution-model, the volume of these edge cells is overestimated and occupies a substantial fraction of the total volume of the cool gas cloud.  This poorly-resolved cloud will quickly evaporate into the hot medium through numerical diffusion and artificial mixing promoted by its overly large warm interface.  
The high-resolution case operates at 8x finer spatial resolution, similar to the enhancement due to EHR in the \textsc{tempest} simulations.  
The smaller grid cells in the high-resolution model more accurately approximate the cloud boundary leading to a longer-lived cool cloud.  We note that this effect will occur for any geometry of cloud to some degree, including filaments and sheets.

We numerically demonstrate this mechanism in Figure \ref{fig:blob} with simple 2D \textsc{enzo} simulations of a cool cloud in pressure equilibrium with a hot halo globally advecting across a fixed grid with periodic boundary conditions (described as the square advection test in \citealt{hopkins:2015}).  The central cloud is cool ($T = 10^{4}$ K) and dense ($\rho = 10^{-26}$ g cm$^{-3}$) relative to its hot ($T = 10^6$ K) and low-density ($\rho = 10^{-28}$ g cm$^{-3}$) surroundings.  We set the size of the cloud to be 4 kpc across, a size only marginally observed for \ion{H}{1}-clouds in our AMR simulations but seen in abundance for our EHR runs (see Section \ref{subsec:size}).   The global velocity of the simulations was set to 150 km/s, the characteristic velocity of cool filamentary inflows in the \textsc{tempest} simulations.  This scenario is simulated in low resolution, using coarse grid cells (1.3 kpc) consistent with our \textsc{tempest} AMR simulations (i.e., 2 physical kpc), and at high resolution, using 8x smaller grid cells (166 pc), similar to the resolutions reached in our simulations employing EHR (i.e., 250 physical pc).  

In reality, such a system should evolve very little over time as it moves across the simulation domain.  However, we observe distinctly different behavior in the low-resolution and high-resolution cases.  We evolve the system for 260 Myr in which the cloud crosses the 40-kpc wide domain (roughly half the freefall time and virial radius of the \textsc{tempest} halo).  The low-resolution cloud exhibits substantial mixing with the ambient medium, whereas little evolution has occurred in the high-resolution model.  These effects are illustrated in the middle and lower rows of Figure \ref{fig:blob} in temperature and \ion{H}{1}-density slices, respectively, showing similarities with our schematic in Figure \ref{fig:schematic_b}.  The total remaining \ion{H}{1} mass is a factor of ten less in the low-resolution case, demonstrating how underresolved clouds quickly evaporate their \ion{H}{1}.  This mechanism properly explains the change in the thermal balance of the CGM in Figure \ref{fig:phase_diff}, such that simulations employing EHR preserve cool, dense gas instead of spuriously converting it to a warm, hot phase through artificial mixing.

Analytic estimates exist parameterizing the timescale over which this numerical diffusion washes out structure $\tau_{\rm diff}$. For a cloud with length $l_{\rm cloud}$ moving at velocity $v$ across a fixed grid with spatial resolution $\Delta x$, the cloud is resolved by $n_{\rm cells} = l_{\rm cloud} / \Delta x$ resolution elements and travels at a normalized velocity $v_{\rm norm} = v / l_{\rm cloud}$.  Its propensity to mix with its surroundings is defined by its numerical diffusivity $\kappa \propto v \Delta x$ with a diffusion timescale $\tau_{\rm diff} \sim l_{\rm cloud}^2 / \kappa$ \citep{toro:2013}.  These calculations indicate coherent clouds will be artificially mixed on shorter timescales when they are poorly resolved and moving quickly: $\tau_{\rm diff} \propto n_{\rm cells} / v_{\rm norm}$.  The analytic estimates derived here are consistent with the observed behavior for poorly-resolved clouds in the low-resolution advection simulations and \textsc{tempest} AMR simulations to quickly evaporate into the hot halo.  

Artificial mixing of cool gas clouds will be further exaggerated by other effects beyond simple numerical diffusion.  Fluid instabilities (e.g., Kelvin-Helmholtz, Rayleigh-Taylor) will amplify resolution effects of artificial mixing beyond the simple diffusion arguments presented here, only making the resolution problems worse in the realistic environment of a cosmological simulation.  Additionally, hydrodynamics codes utilizing lower-order discretization schemes (i.e., piecewise linear reconstruction; PLM) will suffer from increased numerical diffusion.  In this respect, our use of the higher-order piecewise parabolic method (PPM) of spatial reconstruction in \textsc{enzo} partially mitigates these numerical diffusion effects.

Two effects contribute to cool gas clouds being forced to smaller scales.  One, as cool gas flows enter the galactic halo from external sources including filaments and satellite galaxies, the increased pressure in the interior of the galactic halo will compress the cool gas to smaller, more poorly-resolved scales.  Two, as EHR increases the spatial resolution of the halo gas, it not only decreases its numerical diffusion, but it also lowers the numerical viscosity of the gas.  The result is gas with an increased Reynolds number that is more prone to turbulence and fragmentation, which may explain the cascade observed in the \textsc{tempest} simulations to progressively smaller cloud sizes with resolution.

Lastly, we confirm that this proposed mechanism for how EHR prevents cool gas from being transferred to a warm phase is actually operating in the \textsc{tempest} simulations.
In Section \ref{subsec:size} we demonstrated that as the \textsc{tempest} simulations progressively resolved smaller scales in the halo, there was an increase in coherent cool clouds existing on smaller scales.  The median \ion{H}{1} cloud size for each simulation was systematically found to be $\sim$5 cells in its smallest dimension and $\sim$10 cells in its longest dimension, regardless of the absolute spatial scale of the simulation.  This is consistent with the idea that the thermally unstable gas in the halo is ``shattering'' \citep{mccourt:2018} to the smallest scales possible in the simulation.  The smaller cloud sizes found in the \textsc{tempest} simulations employing EHR are evidence that this mechanism is occurring and preserving cool, dense gas against artificial evaporation, reproducing the behavior found in Figure \ref{fig:phase_diff}.

\section{Discussion: Comparisons, Caveats, and Convergence}
\label{sec:caveats}

\subsection{Sources of Cool CGM in the \textsc{tempest} Simulations}

There are four primary sources for cool gas in a galactic halo: (1) filamentary inflows of low-metallicity gas from the IGM; (2) the ISM of accreting satellite galaxies; (3) cool clouds entrained or created in supernova-driven outflows from the disk of the galaxy; and (4) cool gas precipitating out of the halo gas.  Here, we discuss the relative strength of each source in the context of the \textsc{tempest} simulations.  The mechanisms injecting these cool gas sources are all occurring simultaneously in a cosmological simulation, so it can be difficult to identify which ones are active.  However, animations of the time evolution of these systems can help provide assistance in interpreting what is happening\footnote{Movie of galactic \ion{H}{1} evolution is at: \href{http://chummels.org/tempest}{http://chummels.org/tempest}}.  

Upon visually inspecting the \textsc{tempest} simulation movies, we observe that the bulk of the cool gas in the halo arises from external sources, as cool filamentary inflows from the IGM and as stripped ISM of merging galaxies.  Both of these cool gas sources are enhanced through EHR.  Filaments become narrower as they enter the inner galaxy, compressed by pressure, and allowed to efficiently cool with EHR as demonstrated in Figure \ref{fig:filament}.  Alternatively, as accreting galaxies undergo tidal disruption, they fling their cool ISM into extended tidal features, which quickly evaporate in the coarsely-resolved AMR simulation, but fragment, cool, and survive in the higher-resolution EHR simulations.  Consequently, the largest differences between the cool gas CGM content in low- and high-resolution simulations typically occur in the period following a merger with another galaxy.  This may help explain why \citet{vandevoort:2018} find such a large increase in CGM \ion{H}{1} content in their high-resolution simulations, because they analyze a snapshot when a major merger is occurring.

In the \textsc{tempest} simulations, we witness a few instances of supernova-driven winds ejecting cool gas into the halo following large star formation events, but these are a sub-dominant effect relative to external cool gas feeding. The stellar feedback prescription we used in the \textsc{tempest} simulations is a simple model for injecting thermal energy into the cells immediately surrounding a supernova.  Such a model has been demonstrated to be relatively ineffective at generating outflows, so it illustrates the efficacy of EHR in that it creates cool gas outflows in these simulations at all.  Other simulation suites that include more sophisticated stellar feedback models will likely increase the total cool gas content in the CGM due to supernova-driven winds \citep[e.g.,][]{suresh:2018}.

EHR enhances precipitation, but in the \textsc{tempest} simulations we primarily observe it happening only in already slightly overdense cool structures like filaments and with moderate increases in the cool gas content of the halo.  To date, studies of cool gas precipitation out of the galactic halo have largely been confined to idealized simulations lacking a cosmological context, where large quantities of cool gas spontaneously condense in the halo medium as it undergoes runaway cooling \citep{meece:2015, li:2015, voit:2017}. Our work here is among the first to observe precipitation occurring in cosmological simulations, primarily aided by its high-resolution in the halo.  However, our finding that precipitation preferentially occurs in filaments is somewhat in tension with the idealized simulations indicating more widespread precipitation throughout the low-density halo (e.g., Silvia et al., in prep).

This apparent tension in the location and degree to which precipitation occurs in idealized galaxy simulations and cosmological galaxy simulations may be eased by considering the effects of numerical diffusion.  Firstly, recall from our analytic estimates for the diffusion timescale that $\tau_{\rm diff} \propto v_{\rm norm}^{-1}$, indicating that structures moving faster relative to the simulation reference frame (i.e., the hydrodynamical grid) will evaporate more quickly due to numerical diffusion.  Idealized simulations affix a galaxy at the center of the simulation domain, minimizing any motion of gas relative to the grid.  Conversely, cosmological simulations like the \textsc{tempest} simulations follow the evolution of a galaxy as it moves through its environment, being bombarded by rapidly-moving gas flows from nearby galaxies and external filaments.  By allowing the galaxy to move relative to the fixed hydrodynamical grid, cosmological simulations intrinsically amplify the artificial effects of numerical diffusion above those in idealized simulations.  As described in this paper, increasing the effects of numerical diffusion will more efficiently mix slightly overdense, metal-rich, or cool gas structures, effectively removing the seeds necessary for runaway cooling that manifest as precipitation throughout the halo.  Resolution helps to suppress these diffusive effects, but even at the highest level of EHR in the \textsc{tempest} simulations, the only seeds preserved as precipitation sites are in coherent overdensities like filamentary flows and accreting galaxies.

Thus, it seems reasonable to conclude that with higher resolution or higher-order reconstruction methods that more effectively suppress numerical diffusion, cosmological simulations will produce even more widespread precipitation of cool gas out of the galactic halo, consistent with idealized simulations.  Furthermore, the additional inclusion of more sophisticated stellar feedback models coupled with EHR, will launch more cool, metal-enriched gas into the halo as sites for runaway cooling.  For these reasons, we predict that future simulations will observe precipitation to be an increasingly important contributor to the cool gas content of the CGM beyond its moderate effects in the \textsc{tempest} simulations.

\subsection{Comparison with Similar Studies}

There are three other concurrent studies using varying implementations of EHR: \citet{vandevoort:2018}, \citet{peeples:2018}, \citet{suresh:2018}, which we explore here.  \citet{vandevoort:2018} use the \textsc{arepo} moving-mesh hydrodynamics code \citep{springel:2010} to generate cosmological zoom simulations using initial conditions from the \textsc{auriga} project \citep{grand:2017} following a Milky-Way Mass galaxy down to $z=0$.  In addition to the standard moving-mesh treatment of hydrodynamics of these simulations, they develop a method enforcing resolution to a fixed spatial scale in the galactic halo.  Periodically, they run a halo-finder, injecting a ``dye'' scalar field into the halos with $m_{\rm halo} > 10^{\rm 8.7} {\rm M_{\rm \odot}}$, which is advected with the fluid and used to identify regions of additional refinement.  In regions where the dye is $> 90\%$ its original value, a spatial scale is enforced.  They test halo spatial scales of 2 and 1 physical kpc for their target halo and the halos of merging satellites.  In comparison, the \textsc{tempest} simulations investigate 2, 1, and 0.5 physical kpc spatial resolution out and beyond $r_{\rm 200}$, reaching better spatial resolution out to larger distances around the target galaxy.  We study the effects at $z=1$ while they analyze results at $z=0$. 

Like the \textsc{tempest} simulations, \citet{vandevoort:2018} find that increased spatial resolution does not affect the bulk properties of the galaxy, and that it increases the neutral hydrogen column densities.  In apparent conflict with our findings, they show that \ion{H}{1} mass in the CGM does not increase with resolution, and they observe a more pronounced increase in median \ion{H}{1} column density of over a dex throughout much of the halo, whereas we only see an increase of about a factor of two.  On closer inspection, their galaxy is undergoing a major merger at $z=0$, the time of their analysis.  This explains why their CGM \ion{H}{1} mass doesn't appear to change much with resolution, because the ISM of the merging galaxy dominates the \ion{H}{1} budget in the target halo and does not change with resolution.  Furthermore, as discussed above, we see the greatest effects from EHR at periods during and following mergers.  Tidally-disrupted gas from the merging halo is dispersed into the halo and quickly evaporated in the low-resolution case but preserved for high-resolution simulations.  It is possible that the factor of $\sim$5 increase in \ion{H}{1} column densities they observe relative to the \textsc{tempest} simulations is explained by the major merger occurring at $z=0$ in their runs.  However, it is also possible that their more sophisticated stellar feedback model more efficiently produces cool gas winds from the interior of the galaxy, which are then preserved through the EHR mechanisms described herein.  Fundamentally, the \textsc{tempest} simulations and the work of \citet{vandevoort:2018} are generally consistent and confirm that the effects of EHR are not just artifacts of one numerical method.

The \textsc{foggie} simulations \citep{peeples:2018} employ a similar technique to our own, using the same \textsc{enzo} simulation code \citep{bryan:2014} and our \textsc{tempest} initial conditions to follow the evolution of the same L* galaxy.  They focus their analysis at $z=2$, whereas this study examines effects at $z\sim1$.  We both achieve the same spatial resolution of 500 comoving pc out to 100 comoving kpc away from the galaxy; however, they do not employ the nested refinement region technique of EHR, instead having a discontinuous jump at their refinement box boundary of over an order of magnitude in spatial scales, which may lead to numerical artifacts.  We largely analyze different effects of EHR on the behavior of the CGM gas.  While we focus on the thermal balance of the gas and its change in low ion and high ion abundances, they concentrate on the kinematics of ionic absorbers and the impact of EHR on spectral absorption features \citep{peeples:2018} and emission predictions \citep{corlies:2018}.  They do note a moderate increase in the column density of \ion{H}{1} absorbers at $z=2$ in their high-resolution run, generally consistent with our findings at $z=1$ but they find less of an effect than we do.  Since our model included enhanced resolution for a period lasting over twice as long as theirs did and over a larger region around the galaxy, it may explain the enhanced effects of EHR on our simulated galaxy.

Like \citet{vandevoort:2018}, \citet{suresh:2018} use the \textsc{arepo} moving mesh code to simulate the evolution of a massive galaxy with a focus on spatially resolving its CGM.  \citet{suresh:2018} employ a super-lagrangian refinement scheme similar in effects to Enhanced Halo Resolution.  This ``CGM zoom'' refinement technique assigns an effective mass resolution throughout the CGM of 2000 M$_{\odot}$, which yields a median spatial resolution of 95 physical parsecs at $z=2$, about two times higher than the \textsc{tempest} simulations.  They focus their analysis on a $m_{\rm vir}$$\sim$$10^{12}$ M$_{\odot}$ galaxy at $z=2$, a more massive system than the target of the \textsc{tempest} simulations.  \citet{suresh:2018} use the stellar feedback wind prescription that is employed in the \textsc{illustris} simulations \citep{vogelsberger:2013}, a sophisticated model that is in part designed to launch cool gas in outflows.  Despite their impressive resolution, they do not find significant effects from their enhanced halo resolution.  They find only minor changes between their low- and high- resolution simulations for \ion{H}{1} covering fractions, and radial profiles of \ion{Mg}{2} column densities and different phases of CGM gas.  Our studies are in tension, but it may be explained as a product of their feedback prescription effectively ejecting cool gas into the halo at even their ``low'' resolution runs.  As an alternative explanation, the high mass of their target galaxy has a hotter virial temperature leading to less thermally unstable gas in the halo, thereby suppressing the precipitation effects of EHR for physical reasons.

\subsection{Caveats To This Study}

There are a number of limitations to the EHR technique and to the \textsc{tempest} simulations, but we believe that our conclusions are robust despite these caveats.  Here we enumerate each of these caveats and discuss their impact on our findings.

\emph{This is a simulation of a single halo.}  Our results appear promising, but with a single simulated halo, there will always be questions about the veracity of one's findings.  We have made every effort to measure the conditions in these simulations over a range of snapshots and from different projection angles to wash out any temporal or spatial artifacts specific to this simulation.  It is absolutely imperative to test the techniques and the conclusions we have reached in this study with multiple halos and simulation codes, and our future work includes this.  Notably, many of the results here are borne out by other groups using similar techniques at different redshifts \citep[e.g.,][]{vandevoort:2018, peeples:2018} suggesting our results are robust.

\emph{The neutral hydrogen content for the \textsc{tempest} simulations with EHR still falls short of observational constraints.}  The EHR technique described in this paper attempts to address a current failure of traditional simulations used to model the CGM, but we make no claims that this is the only effect responsible for an underabundance of cool gas modeled in simulated galactic halos.  There are many other avenues for potentially increasing the cool gas content of the CGM, such as including B-fields \citep{ji:2018a}, cosmic rays \citep{salem:2016, butsky:2018}, or more sophisticated feedback techniques.  Alternatively, it may simply be that we have not achieved high enough spatial resolutions since our behavior has not yet converged.  Future generations of the \textsc{tempest} simulations will investigate the combined role of additional physics with resolution in the halo in modifying the properties of the CGM.

\emph{The \textsc{tempest} simulations use overly simplistic physics.}  We chose to use relatively simple physics for this first-generation of \textsc{tempest} simulations to isolate and investigate the effects of spatial resolution in the halo.  Specifically, 
the stellar feedback prescription is a thermal-only model, depositing supernovae energy as thermal energy distributed over the 3x3x3 cells centered on the young stellar particle.  It has been shown in numerous studies that such a simple thermal-only feedback prescription is unable to reproduce all the observational characteristics of galaxies \citep[e.g.,][]{steinmetz:1999, hummels:2012}.  Furthermore, these simulations do not include effects of self-shielding \citep[e.g.,][]{rahmati:2013} on the treatment of gas ionization states and cooling, though hydrogen and helium ions are followed in their full non-equilibrium evolution.  However, despite the lack of more realistic stellar feedback and cooling, the \textsc{tempest} simulations still demonstrate the trend of increasing CGM cool gas content with EHR, and it is likely that the inclusion of both of these effects will only amplify the resulting cool gas and low ion content.

\subsection{Qualitative Trends and Predictions for a Misty CGM}

The primary result of this study is that increasing the spatial resolution in the galactic halo improves the modeling of the halo gas and produces a number of physical and observable changes in the CGM.  We find no evidence for convergence in these CGM properties based on the \textsc{tempest} simulations.  But we can make predictions about the true nature of the CGM based on the trends we observe with Enhanced Halo Resolution.

By resolving progressively smaller scales in cosmological simulations of the CGM, we demonstrate a shift in its thermal balance, enhancing cool gas content at the expense of warm gas.  This cool gas is found in an increasing number of small, low-mass clouds at the resolution scale of the simulation, able to survive as coherent structures for progressively longer periods before evaporating into the surrounding hot halo.  The primary prediction for the scale at which these cool gas clouds finally stop their fragmentation is the ``shattering'' sub-parsec lengthscale \citep{mccourt:2018}.  Thus, the trends found in this study are consistent with a misty circumgalactic medium, one composed of ubiquitous sub-parsec cool cloudlets entrained in a hot halo much like a terrestrial fog \citep[e.g.,][]{liang:2018b}.  Observationally, the enhanced cool gas content boosts the \ion{H}{1} and low ion content of the CGM, relieving the current tension between simulations and observations, but it leads to a drop in high ion content somewhat at odds with current observational constraints.  As previously mentioned, additional physical effects ignored in these simulations (e.g., AGN feedback, cosmic rays) may alleviate this tension. 

The source of the cool gas in the \textsc{tempest} simulations is primarily external due to IGM accretion and the ISM of merging galaxies, but there is reason to believe that with more realistic stellar feedback and even higher resolution, the amount of cool gas contributed from internal sources like galactic outflows and thermal precipitation will increase.  Thus, we predict that all four sources remain viable for contributing to the true cool gas content of the galactic halo. 

Additional simulations using other codes, more physics, and higher resolution will be necessary to confirm the trends we propose here.  However, the method of EHR is computationally intensive, and it will be increasingly challenging to extend this prescription to sub-parsec scales using current simulation technologies.  There is promising work being performed using the GPU-optimized hydro code \textsc{cholla} \citep{schneider:2015} to self-consistently model spatial scales of 5~pc out to galactic radii of 10 kpc \citep{schneider:2018a} ($10^{10}$ resolution elements) using the entirety of the Department of Energy's Titan supercomputer.  This state-of-the-art effort is approaching our current computational limits.  A quick calculation shows that to tile 1-parsec resolution elements filling an L* galactic halo ($r_{\rm vir}\sim250$ kpc) requires $10^{17}$ resolution elements.  Barring any significant algorithmic improvements, Moore's Law suggests that we need $\sim$35 years before computers will be efficient enough to perform this sort of simulation at ``shattering'' resolution scales.  Therefore, alternative approaches for modeling the small-scale behavior of the CGM at extremely high resolutions will potentially turn to subgrid models to represent unresolved gas dynamics and composition.

\subsection{EHR for Particle-Based Codes}

Our description of EHR is predicated on the use of a grid-based simulation code (e.g., \textsc{enzo}, \textsc{art}, \textsc{ramses}, \textsc{athena}) because our use of discrete regions to force a minimum spatial resolution is intrinsically tied to an Eulerian implementation of hydrodynamics.  This description is limiting because there are a number of additional codes used to simulate galaxy evolution employing other methods, namely particle-based methods like smoothed particle hydrodynamics (SPH).  

Because they are a fundamentally different manner of representing fluids, Lagrangian codes (i.e., particle-based) have a different set of limitations than grid-based codes.  When cloud sizes approach the resolution scale in an Eulerian code, the simulation tends to overestimate the mixing of that cloud with its surroundings, whereas Lagrangian codes generally suppress fluid mixing, resulting in a numerically-induced surface tension and the artificial preservation of small structures \citep[e.g.][]{agertz:2007}.  Thus, the simulations still break down at small resolution scales but in a different way, so they could still benefit from a technique like EHR.

However, generalizing our method of EHR to work with a particle-based code is challenging, because of how spatial resolution is calculated in a Lagrangian representation of hydrodynamics.  The spatial resolution in a particle-based code is tied to the gas smoothing length, directly related to the particle density.  Thus, to force additional spatial resolution in the halo of a particle-based simulation, one must increase the particle density in the low-density halo, potentially by particle-splitting methods to break particles into more numerous, lower-mass particles in the halo \citep[e.g.,][]{kitsionas:2012}.  Unless confined, these low-mass particles will drift into regions in the simulation (e.g., the disk) where more massive particles are present.  Particles of different masses interacting gravitationally can lead to unphysical effects like numerical scattering as the system seeks energy equilibrium and increases the velocity of low-mass particles.  

Therefore, the prospect of implementing EHR for a particle-based simulation code is a great challenge requiring diligent bookkeeping to assure different regions of the simulation only contain particles of a single mass through particle splitting and merging methods.

\section{Conclusions}
\label{sec:conclusions}

Our results can be summarized as:

\begin{enumerate} 

\item We describe Enhanced Halo Resolution, a novel technique for improving the hydrodynamical modeling of the galactic halo.  In general, EHR is any technique for maintaining a base-level resolution in a region centered on the galaxy and extending out into the galactic halo and beyond.  In our implementation, EHR employs a series of nested regions to assure that resolution elements never drop below a specified fixed physical scale throughout the redshift evolution of the simulation.  EHR can be combined with traditional AMR to allow additional density-based refinement on dense gas structures beyond the base-level resolution provided by EHR.

\item We introduce the \textsc{tempest} simulations, a set of grid-based cosmological hydrodynamics zoom simulations that achieve the highest in situ spatial resolution for the CGM to date (500 comoving pc spatial resolution for $r_{\rm gal} < $100 comoving kpc) at $z<2$.  They achieve this unprecedented resolution through use of the EHR technique combined with standard density-based AMR.  The resulting simulated galaxy halos provide a test bed for demonstrating the effects of EHR and making increasingly accurate models of the CGM.

\item We demonstrate with the \textsc{tempest} simulations how EHR changes the CGM by (1) more continuously and more correctly sampling the various properties of the CGM (e.g., density, temperature, metallicity); (2) increases its cool gas content and observed column densities of low ions (e.g., \ion{H}{1}); (3) slightly reduces CGM warm gas content and corresponding column densities of the high ions (e.g., \ion{O}{6}) probing that gas; and (4) progressively decreases the size and mass of \ion{H}{1}-bearing clouds to the resolution scale of the simulation.  Notably, through these effects, EHR provides a means of breaking the longstanding problem of simulations underpredicting the observed low ion covering fractions and column densities of the CGM.

\item We suggest two mechanisms for explaining why EHR produces the observed effects on the CGM: (1) Increased spatial resolution more accurately samples gas properties, including its extremes in density, temperature, and metallicity where the cooling rate is substantially higher.  These extremes act as seed sites for runaway cooling and precipitation of cool gas out of the warm/hot CGM.  (2) Additional resolution minimizes numerical diffusion on poorly resolved structures, thus preventing extant cool gas sources from evaporating due to unphysical mixing with the surrounding hot medium.  We confirm that both mechanisms are occurring in the \textsc{tempest} simulations and provide animations illustrating these effects.

\item We enumerate the various sources for cool gas content in the CGM including: (1) filamentary inflows of pristine gas from the IGM; (2) the ISM of accreting galaxies; (3) supernova-driven outflows from the galaxy itself; and (4) precipitation from the ambient hot halo.  We describe how the two mechanisms of EHR can enhance any of these sources of cool gas, but the \textsc{tempest} simulations are dominated by external sources (1 \& 2) likely due to ineffective thermal supernova feedback.  Concurrent studies employing other EHR implementations show enhanced cool gas content \citep{vandevoort:2018} from winds \citep{suresh:2018} probably due to EHR additionally enhancing cool gas expelled as outflows driven by more effective supernova feedback prescriptions.

\item Using analytic estimates of the diffusion timescale, we offer up an explanation for why cosmological simulations suppress the development of precipitation beyond idealized galaxy simulations.  In traditional cosmological simulations, both the coarse resolution as well as the galaxy's motion relative to its hydrodynamic grid, amplify its numerical diffusivity, evaporating most of the seed structures necessary for the growth of runaway cooling.  In the \textsc{tempest} simulations including EHR, we address the coarse resolution aspect, enabling precipitation to occur in already coherent overdense structures like filamentary inflows.  We predict that as resolution continues to improve, simulations will find evidence for increased levels of precipitation throughout the rest of the halo.  This effect will be further amplified by the inclusion of more sophisticated feedback models that more efficiently drive cool, metal-enriched gas into the halo to act as seeds for precipitation.

\item We find no convergence in the effects of EHR on the CGM, so we extrapolate on the trends of increasing resolution to make qualitative predictions on the nature of the CGM in reality (i.e., at infinitely-high resolution).

\emph{We predict that the CGM is a mist, consisting of a large number of small (sub-pc), low-mass (sub-M$_{\odot}$), long-lived, cool ($T\sim10^4$K) cloudlets entrained in a hot medium at the virial temperature of the galactic halo, arising from a number of different sources both internally and externally.}  We further predict that because of computational challenges, future efforts to reach convergent behavior for the CGM will increasingly rely upon subgrid models to represent the dynamics and composition of halo gas on sub-parsec scales.
\end{enumerate}

\acknowledgments

The authors are very grateful to X. Prochaska, Kate Rubin, Greg Bryan, Shea Garrison-Kimmel, Claude-Andr\'{e} Faucher-Gigu\`{e}re, and Anatoly Klypin for helpful discussions.  In particular, we thank Matt Turk, Nathan Goldbaum, John Zuhone, Kacper Kowalik, Bili Dong and other members of our \textsc{yt} developer community, who were invaluable in producing and maintaining much of the code necessary to process these datasets.  Support for CBH and NL was provided by NASA through Hubble Space Telescope (HST) theory grants HST-AR-13917, HST-AR-13919 with additional funding for CBH and DS from the National Science Foundation (NSF) Astronomy and Astrophysics Postdoctoral Fellowship program. BDS was supported by NSF grant AST-1615848.  Support for PFH was provided by an Alfred P. Sloan Research Fellowship, NSF Collaborative Research Grant \#1715847 and CAREER grant \#1455342, and NASA grants NNX15AT06G, JPL 1589742, 17-ATP17-0214.  BWO was funded in part by NSF grants PHY-1430152, AST-1514700, AST-1517908, OAC-1835213, by NASA grants  NNX12AC98G, NNX15AP39G, and by HST-AR-14315.  JHW was supported by NSF grants AST-1614333 and OAC-1835213, NASA grant NNX17AG23G, and HST theory grant HST-AR-14326.  ISB received partial support from HST-AR-15046.

These simulations were performed and analyzed on Blue Waters, operated by the National Center for Supercomputing Applications (NCSA) with PRAC allocation support by the NSF (award number ACI-1514580) and the Great Lakes Consortium for Petascale Computation. This research is part of the Blue Waters sustained-petascale computing project, which is supported by the NSF (awards OCI-0725070, ACI-1238993, and ACI-1514580) and the state of Illinois. Blue Waters is a joint effort of the University of Illinois at Urbana-Champaign and its NCSA.  Additional computational resources for this work from from NSF XSEDE resources under allocation TG-AST140018.  Computations described in this work were performed using the publicly-available \textsc{enzo} code, which is the product of a collaborative effort of many independent scientists from numerous institutions around the world.

\software{
\textsc{enzo} \citep{bryan:2014},
\textsc{matplotlib} \citep{hunter:2007},
\textsc{numpy} \citep{vanderwalt:2011},
\textsc{trident} \citep{hummels:2017},
\textsc{yt} \citep{turk:2011},
\textsc{yt} \textsc{astro} \textsc{analysis} \citep{smith:2018b}
\textsc{ytree} \citep{smith:2018a}
}

\bibliography{main}

\end{document}